\documentclass[12pt]{article}
\usepackage[dvips]{graphicx}
\usepackage{epsfig}
\usepackage{latexsym}
\usepackage{amssymb}
\usepackage{amsmath}

\usepackage{wrapfig}
\usepackage{boxedminipage}
\usepackage{setspace}
\usepackage{subfigure}
\DeclareSymbolFont{AMSb}{U}{msb}{m}{n}
\DeclareMathSymbol{\IN}{\mathbin}{AMSb}{"4E}
\DeclareMathSymbol{\IZ}{\mathbin}{AMSb}{"5A}
\DeclareMathSymbol{\IR}{\mathbin}{AMSb}{"52}
\DeclareMathSymbol{\Q}{\mathbin}{AMSb}{"51}
\DeclareMathSymbol{\II}{\mathbin}{AMSb}{"49}
\DeclareMathSymbol{\IC}{\mathbin}{AMSb}{"43}
\DeclareMathSymbol{\IP}{\mathbin}{AMSb}{"50}
\DeclareMathSymbol{\IH}{\mathbin}{AMSb}{"48}
\DeclareMathSymbol\IA{\mathalpha}{AMSb}{"41}
\DeclareMathSymbol\IS{\mathalpha}{AMSb}{"53}

\def\Q{{\cal Q}}

\begin{document}

\begin{flushright}
\phantom{{\tt arXiv:0802.????}}
\end{flushright}

\bigskip
\bigskip
\bigskip

\begin{center} {\Large \bf  External Fields  and Chiral Symmetry Breaking}
  
  \bigskip

{\Large\bf  in the}

\bigskip

{\Large\bf  Sakai--Sugimoto Model}

\end{center}

\bigskip \bigskip \bigskip \bigskip

\centerline{\bf Clifford V. Johnson, Arnab Kundu}

\bigskip
\bigskip

  \centerline{\it Department of Physics and Astronomy }
\centerline{\it University of
Southern California}
\centerline{\it Los Angeles, CA 90089-0484, U.S.A.}

\bigskip

\centerline{\small \tt johnson1, akundu@usc.edu}

\bigskip
\bigskip


\begin{abstract} 
\noindent 
Using the Sakai--Sugimoto model we study the effect of an external
magnetic field on the dynamics of fundamental flavours in both the
confined and deconfined phases of a large $N_c$ gauge theory. We find
that an external magnetic field promotes chiral symmetry breaking, consistent
with the ``magnetic catalysis'' observed in the field theory
literature, and seen in other studies using holographic duals. The
external field increases the separation between the deconfinement
temperature and the chiral symmetry restoring temperature. In the
deconfined phase we investigate the temperature-magnetic field phase
diagram and observe, for example, there exists a maximum critical
temperature (at which symmetry is restored) for very large magnetic
field. We find that this and certain other phenomena persist for the
Sakai--Sugimoto type models with probe branes of diverse dimensions.
We comment briefly on the dynamics in the presence of an external
electric field.

\end{abstract}
\newpage \baselineskip=18pt \setcounter{footnote}{0}

\section{Introduction}

Since the advent of the AdS/CFT
correspondence\cite{Maldacena:1997re,Witten:1998zw,Gubser:1998bc},
(see a review in ref.\cite{Aharony:1999ti}) there has been
considerable refinement of the methods for studying strongly coupled
large $N_c$ gauge theories. Much effort has been spent studying such systems at both
zero and finite temperature, constructing specific ``holographic''
models intended to capture key features of QCD at strong coupling,
such as the confinement/deconfinement phase transition, chiral
symmetry breaking, and possible novel phases that may be of relevance
to experiment and observation.

The Sakai--Sugimoto model, as described in ref.~\cite{Sakai:2004cn} is
one such construction which cleanly realizes chiral symmetry breaking
and deconfinement. The supergravity background of this model is
constructed of near-horizon geometry of $N_c$ $D4$-branes, following
ref.\cite{Witten:1998zw}. The study of $N_f$ flavour $D8$ branes in
this background when $N_f\ll N_c$ reveals a nice geometric realization
of chiral symmetry breaking. The flavour branes do not backreact on
the background geometry in this probe limit and therefore studying
their dynamics using the Dirac-Born-Infeld (DBI) action (including a
Wess-Zumino term, if necessary) suffices to capture the corresponding
gauge theory dynamics of fundamental flavours in an analogue of the
quenched approximation.

In this (DBI) regime it is possible to capture general gauge theory
features such as the phase diagram for temperature vs chemical
potential by considering probe brane in finite temperature
supergravity background and exciting specific gauge field on the
world--volume of the probe brane. Previous such studies including the
non--zero chemical potential in this model have been carried out in
e.g.,
refs.~\cite{Horigome:2006xu,Rozali:2007rx,Bergman:2007wp,Davis:2007ka}.
Here, we will introduce an external magnetic and electric field.

A clear method for introducing a background magnetic field has been
previously discussed in the $D3/D7$ model in ref.~\cite{Filev:2007gb}.
The authors consider pure gauge $B$--field in the supergravity
background, which is equivalent to exciting a gauge field on the
world-volume of the flavour branes corresponding to a magnetic field.

We find that the presence of magnetic field promotes the spontaneous
breaking of chiral symmetry. This is expected from the field theory
perspective and is widely recognized as a sort of ``magnetic
catalysis'' for chiral symmetry breaking (see e.g.,
ref.~\cite{Miransky:2002eb}). Further study of the phase structure of
this model reveals the existence of a finite critical temperature (for
restoration of chiral symmetry) for large magnetic field. We
 analyze a number of other physical quantities  such as
 the latent heat and relative magnetisation associated to the phase
transition. 

We also find that our phase structure is rather generic for the
Sakai--Sugimoto type holographic models where we consider the dynamics
of probe $Dp$-brane in $D4$-brane background. We address some of the
physics of an external electric field in the model. We find that in
the symmetry--restored phase an external electric field drives a
current in the gauge theory due to pair creation, and the symmetry--broken
phase does not conduct. However we have not considered the presence of baryons in our set-up, which could give rise to a non-zero current in the phase where chiral symmetry is broken.

This paper, is organised as follows: We briefly review the
Sakai--Sugimoto construction in section 2. In section 3 we perform the
analysis of probe $D8$--branes in the presence of magnetic field, while in section
4 we discuss similar results for general probe $Dp$-brane. In section
5 we comment on some aspects of the physics of an external electric
field, concluding in section 6.

{\bf Note added:} When this paper was being prepared we became aware
of ref.~\cite{Bergman:2008sg}, in which authors have studied related
physics.

\section{The Sakai--Sugimoto Construction}

The Sakai--Sugimoto model\cite{Sakai:2004cn} consists of near-horizon limit
of $N_c$ $D4$-branes wrapped on a circle of radius $R$ in the $x^4$
direction with anti-periodic boundary condition for fermions. The
$D4$-branes are intersected in the compact $x^4$ direction by $N_f$
$\overline{D8}$-branes at $x^4=-\frac{L}{2}$ and $N_f$ $D8$-branes at
$x^4=\frac{L}{2}$ (with the constraint that $L\le \pi R$). This is
dual to a $(4+1)$-dimensional $SU(N_c)$ Yang-Mills theory with gauge
coupling constant $g_5$; left and right handed quarks are introduced
by the flavour $\overline{D8}$ and $D8$-branes in the probe limit that share three spatial directions with the $D4$-branes. The
flavour branes introduce a global flavour symmetry $U(N_f)_L\times
U(N_f)_R$ as seen from the $(4+1)$-dimensional $D4$-brane worldvolume
gauge theory. This global symmetry is identified with the chiral symmetry (non-abelian) of the effective $(3+1)$-dimensional gauge theory with chiral fermions. In the probe limit, the dynamics of the flavour branes
is described by the DBI action in the background of $N_c$ $D4$-brane
geometry. The background metric of $D4$-brane is obtained from type
IIA supergravity and is given by
\begin{eqnarray}\label{eqt: metsakai}
&& ds^2 = \left(\frac{u}{R_{D4}}\right)^{3/2}\left(-dt^2+dx_idx^i+f(u)(dx^4)^2\right)+\left(\frac{u}{R_{D4}}\right)^{-3/2}\left(\frac{du^2}{f(u)}+u^2d\Omega_4^2\right)\ ,   \nonumber\\
&&   e^{\phi}=g_s\left(\frac{u}{R_{D4}}\right)^{3/4}\ , \quad F_{(4)}=\frac{2\pi N_c}{V_4}\epsilon_4\ ,\quad f(u)=1-\left(\frac{U_{KK}}{u}\right)^3\ .
\end{eqnarray}
Here $x^i$ are the flat $3$-directions, $t$ is the time coordinate,
$x^4$ is the spatial compact circle, $\Omega_4$ are the $S^4$
directions and $u$ is the radial direction. $l_s$ is the string
length, $g_s$ is the string coupling; $V_4$ and $\epsilon_4$ are the
volume and volume form of $S^4$ respectively. Also, $\phi$ is the
dilaton and $F_{(4)}$ is the RR four-form field strength. To avoid a
conical singularity in the $\{x^4,u\}$ plane one should make periodic
identification:
\begin{equation}
\delta x^4=\frac{4\pi}{3}\left(\frac{R_{D4}^3}{U_{KK}}\right)^{1/2}=2\pi R\ .
\end{equation}

This endows the background with a smooth cigar geometry in the $\{x^4,u\}$ plane. The radial parameter $R_{D4}^3$ is given by
\begin{equation}
R_{D4}^3=\pi g_sN_cl_s^3=\pi\lambda\alpha'\ ,
\end{equation}
where $\lambda$ is the 't-Hooft coupling.  This construction has three
dimensionful parameters $g_5$, $L$ and $R$, where the $L$ and $R$ have
been defined above. The five-dimensional gauge coupling is given by
$g_5^2=(2\pi)^2g_sl_s$. The four dimensional gauge coupling can be
obtained by dimensional reduction yielding $g_4^2=g_5^2/2\pi R$. The
five dimensional 'tHooft coupling is defined to be $\lambda=(g_5^2
N_c)/4\pi$. The gravity picture is valid for small curvature which
amounts to $\lambda\gg R$, namely for strong 'tHooft coupling. It should be noted that due to the presence of a varying dilaton in equation (\ref{eqt: metsakai}) the type IIA supergravity background becomes unreliable in the far UV and we need to lift it to M-theory for a possible UV-completion.

To consider the finite temperature version of the model we need to
Euclideanise the background in equation~(\ref{eqt: metsakai}). This
can be achieved by compactifying the time direction, t, on a circle
and identifying the period with inverse temperature $\beta$. In this
case, the $x^4$ circle shrinks away at $u=U_T$ but the $t$ circle is
fixed. One can also construct a finite temperature version by
interchanging the role of the $t$ and $x^4$ circles so that now time
circle shrinks away at some value $u=U_T$ but the $x^4$ circle remains
fixed. It is easy to see that both these constructions have the same
asymptotic behaviour. These are the only known Euclidean continuations
of the background in equation~(\ref{eqt: metsakai}) with the right
asymptote. It is known (e.g, in ref.~\cite{Aharony:2006da}) that for
$T<1/2\pi R$ (i.e low temperature), the background with the $x^4$
circle shrinking dominates, where for $T>1/2\pi R$ (i.e., high
temperature) the background with the $t$ circle shrinking dominates;
and this geometric transition between the two background corresponds
to the confinement/deconfinement transition.

 So for low temperature the relevant background is given by equation~(\ref{eqt: metsakai}) with the time coordinate periodically identified with period $\beta=1/T$, where $T$ is the temperature. The high temperature background is given by
\begin{eqnarray}\label{eqt: highmet}
&& ds^2=\left(\frac{u}{R}\right)^{3/2}\left(dx_idx^i+f(u)dt^2+(dx^4)^2\right)+\left(\frac{u}{R}\right)^{-3/2}\left(\frac{du^2}{f(u)}+u^2d\Omega_4^2\right)\ ,\nonumber\\
&& t=t+\frac{4\pi R_{D4}^{3/2}}{3U_T^{1/2}}\ ,\quad T=\frac{1}{\beta}=\left(\frac{4\pi R_{D4}^{3/2}}{3U_T^{1/2}}\right)^{-1}\ ,\quad f(u)=1-\left(\frac{U_T}{u}\right)^3\ .\\ \nonumber
\end{eqnarray}
All the parameters are given by the same formula as equation~(\ref{eqt: metsakai}). The dilaton, RR $4$-form, $R_{D4}$ are also given by the same formula as equation~(\ref{eqt: metsakai}).
Now one can introduce the flavour brane--anti-brane system in the probe limit, namely $N_f\ll N_c$. In this limit the probe branes do not backreact on the geometry and the classical profile of the probe is solely determined by the Dirac-Born-Infeld action. We will consider the following ansatz for the flavour $\overline{D8}$-$D8$ branes:
\begin{equation}\label{eqt: ansatz}
\{t,x_i,x^4=\tau,\Omega_4,u=u(\tau)\}\ .
\end{equation}

For notational convenience we rename $x^4$ coordinate to be $\tau$, and we note that the coordinates in the parenthesis should be understood as the worldvolume coordinates of the $\overline{D8}/D8$-brane.

\section{The Probe Brane Analysis}

Many aspects of the finite temperature physics have been studied
before, e.g. in refs.~\cite{Aharony:2006da},\cite{Parnachev:2006dn}
and references therein. To introduce external magnetic field we follow
the procedure adopted in ref.~\cite{Filev:2007gb}. We consider the
presence of a pure gauge $B$-field given by, $B_2=Hdx^2\wedge dx^3$.
As far as the DBI action is concerned it can be easily seen that such
a choice is equivalent to exciting a gauge field $A_3=H x^2$ on the
worldvolume of the probe brane.

\subsection{The Low Temperature Background}

In this case the relevant background is given by equation~(\ref{eqt:
  metsakai}). With the probe brane ansatz in equation~(\ref{eqt:
  ansatz}) the induced metric on the worldvolume of
$\overline{D8}/D8$-brane is given by
\begin{eqnarray}\label{eqt: indlowtem}
ds_{D8}^2 &=& \left(\frac{u}{R_{D4}}\right)^{3/2}\left(dt^2+dx_idx^i\right)+\left(\frac{R_{D4}}{u}\right)^{3/2}\left(f(u)+u'^2\left(\frac{R_{D4}}{u}\right)^3\frac{1}{f(u)}\right)d\tau^2 \nonumber\\
&& +\left(\frac{u}{R_{D4}}\right)^{-3/2}u^2d\Omega_4^2\ ,\\ \nonumber
\end{eqnarray}
where $u'=du/d\tau$; also we drop the negative sign in front of the
time coordinate as we are considering the Euclidean metric. The DBI
action is given by\footnote{It is a straightforward exercise to check that in this case the Wess-Zumino term does not contribute to the probe brane action.}
\begin{eqnarray}\label{eqt: dbi}
S_{D8} &=& \mu_8\int d^9\xi e^{-\phi}\sqrt{det(P[G_{\mu\nu}+B_{\mu\nu}])}
     = C\int d\tau \mathcal{L}(u,u')\ ,
\end{eqnarray}
where $C=\mu_8V_{S^4}V_{R^3}/g_sT$. Recall that we put the flavour branes ($\overline{D8}/D8$) with the asymptotic condition that as $u\to \infty$, $\tau\to\pm L/2$; where the $\pm$ corresponds to the $\overline{D8}$ and $D8$-brane respectively. Also note that the coordinate $\tau$ is restricted from $-\pi R$ to $+\pi R$.

One can immediately note from equation~(\ref{eqt: dbi}) that $\mathcal{L}(u,u')$ is independent of $\tau$, therefore the Hamiltonian corresponding to $\tau$ will be a constant of motion. Carrying out the following Legendre transformation we should have
\begin{equation}\label{eqt: hamil}
\mathcal{H}_{\tau}=u'\frac{\partial\mathcal{L}(u,u')}{\partial u'}-\mathcal{L}(u,u')={\rm const.}\\
\end{equation}
So the first integral of motion that follows from equation~(\ref{eqt: hamil}) is given by
\begin{equation}\label{eqt: firstint}
u^4\frac{\left(1+H^2\left(\frac{R_{D4}}{u}\right)^3\right)^{\frac{1}{2}}f(u)}{\left(f(u)+\left(\frac{R_{D4}}{u}\right)^3\frac{u'^2}{f(u)}\right)^{\frac{1}{2}}}=U_0^4\left(1+H^2\left(\frac{R_{D4}}{U_0}\right)^3\right)^{\frac{1}{2}}\sqrt{f(U_0)}\ .\\
\end{equation}

We have rewritten the constant in the right hand side in a convenient
way. Note that $U_0$ is the minimum value of $u$ that the probe brane
can reach satisfying $u'|_{u=U_0}=0$\footnote{We assume that the brane--anti-brane pair join smoothly, which implies that there is no resultant force present at the point where they meet. Typically this would mean that there is no other source (e.g., a baryon vertex or a bunch of F-strings) present at this point.}. For zero background magnetic
field this set up reduces to the low temperature case analyzed in
ref.~\cite{Aharony:2006da}. Let us now focus on the solution for the
probe brane profile.

We will compare the behaviour of the brane profile in the presence of
magnetic field to the case when it is turned off. For notational
convenience we define the following:
\begin{eqnarray}\label{eqt: rescale}
y=\frac{u}{U_0}\ , \quad y_{KK}=\frac{U_{KK}}{U_0}\ , \quad R_{D4}=U_0d\ , \quad L=U_0l\ . \nonumber
\end{eqnarray}
With the above redefinitions we can obtain the difference in slope of the profile in presence and in absence of magnetic field as
\begin{eqnarray}\label{eqt: diffslope}
u_H'^2-u_{H=0}'^2=f(y)^2\frac{f(y)}{f(1)}\left(\frac{y}{d}\right)^3y^8\frac{H^2d^3}{1+H^2d^3}\left(\frac{1}{y^3}-1\right)\ , \nonumber
\end{eqnarray}
with $y\in [1,\infty]$. So we get that $|u_H'|\le |u_{H=0}'|$ for each value of $y$. This in turn means that the magnetic field bends the profile of the $D8/\overline{D8}$ brane and therefore forces the brane--anti-brane pair to join closer to the boundary (and hence break chiral symmetry) for fixed asymptotic separation. 

We can study this explicitly as follows. The brane--anti-brane separation at the boundary ($u\to~\infty$) is given by
\begin{eqnarray}\label{eqt: sepmag}
\frac{L}{2}=\int d\tau =\int_{U_{0(H)}}^\infty\frac{du_H}{u_H'}
           &=&\frac{R_{D4}^{3/2}}{\sqrt{U_{0(H)}}}\int_1^\infty \frac{y^{-3/2}dy}{f(y)\left[\frac{1+H^2\left(\frac{d}{y}\right)^3}{1+H^2d^3}\frac{f(y)}{f(1)}y^8-1\right]^{1/2}}\nonumber\\
           &=&\int_1^\infty \mathcal{I}_{(H)}(y)dy\ .
\end{eqnarray}
Clearly putting $H=0$ we get the corresponding separation when the background magnetic field is switched off.
\begin{eqnarray}\label{eqt: sepnomag}
\frac{L}{2}=\frac{R_{D4}^{3/2}}{\sqrt{U_0}}\int_1^\infty\frac{y^{-3/2}dy}{f(y)\left(\frac{f(y)}{f(1)}y^8-1\right)^{1/2}}
           =\int_1^\infty dy \mathcal{I}_{H=0}\ .
\end{eqnarray}

\begin{figure}[!ht]
\begin{center}
\includegraphics[angle=0,
width=0.65\textwidth]{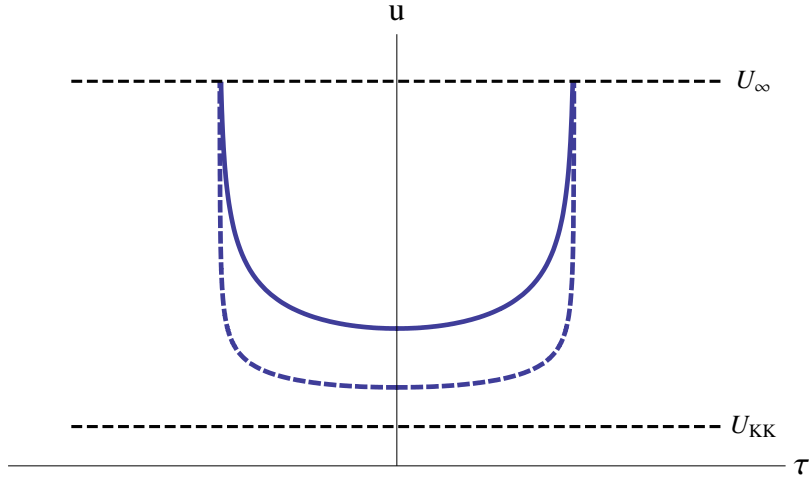}
\caption{\small The dashed U-shaped curve represents a profile in vanishing background field and the solid U-shaped curve represents a profile when a non-zero magnetic field is present. These profiles are obtained by numerically solving the equation of motion for the probe brane.}
\label{fig: plow}
\end{center}
\end{figure}

For the same asymptotic separation magnetic field changes the brane profile's point of closest approach $U_{0(H)}$. We can compare $U_{0(H)}$ and $U_{0}$. Equating equation~(\ref{eqt: sepmag}) and equation~(\ref{eqt: sepnomag}) one gets
\begin{equation}\label{eqt: compa}
\sqrt{\frac{U_{0(H)}}{U_{0}}}=\frac{\int_1^\infty \mathcal{I}_{H=0}dy}{\int_1^\infty \mathcal{I}_{(H)}dy}\ .\\
\end{equation}\\
Some algebra shows that $\mathcal{I}_{H=0}\ge \mathcal{I}_{(H)}$ for all $y$, so the ratio on the right hand side is greater than or equal to one, which also means that $U_{0(H)}\ge U_{0}$. Therefore for the same asymptotic separation the magnetic field can only help to join the brane--anti-brane pair favouring chiral symmetry breaking. This is pictorially represented in figure \ref{fig: plow}.

We can extract more qualitative features in appropriate limits. To do so, let us rewrite equation~(\ref{eqt: sepmag}) with the change to variable $z=y^{-3}$. With this equation~(\ref{eqt: sepmag}) becomes
\begin{eqnarray}\label{eqt: sepz}
\frac{L}{2}=\frac{R_{D4}^{3/2}}{3\sqrt{U_{0}}}\sqrt{\left(1-y_{KK}^3\right)\left(1+H^2d^3\right)}\int_0^1 \frac{\left(1-y_{KK}^3z\right)^{-1} z^{-5/6}dz}{\sqrt{\frac{\left(1-y_{KK}^3z\right)\left(1+H^2d^3z\right)}{z^{8/3}}-\left(1-y_{KK}^3\right)\left(1+H^2d^3\right)}}\ .
\end{eqnarray}

\begin{figure}[h!]
\begin{center}
\subfigure[] {\includegraphics[angle=0,
width=0.45\textwidth]{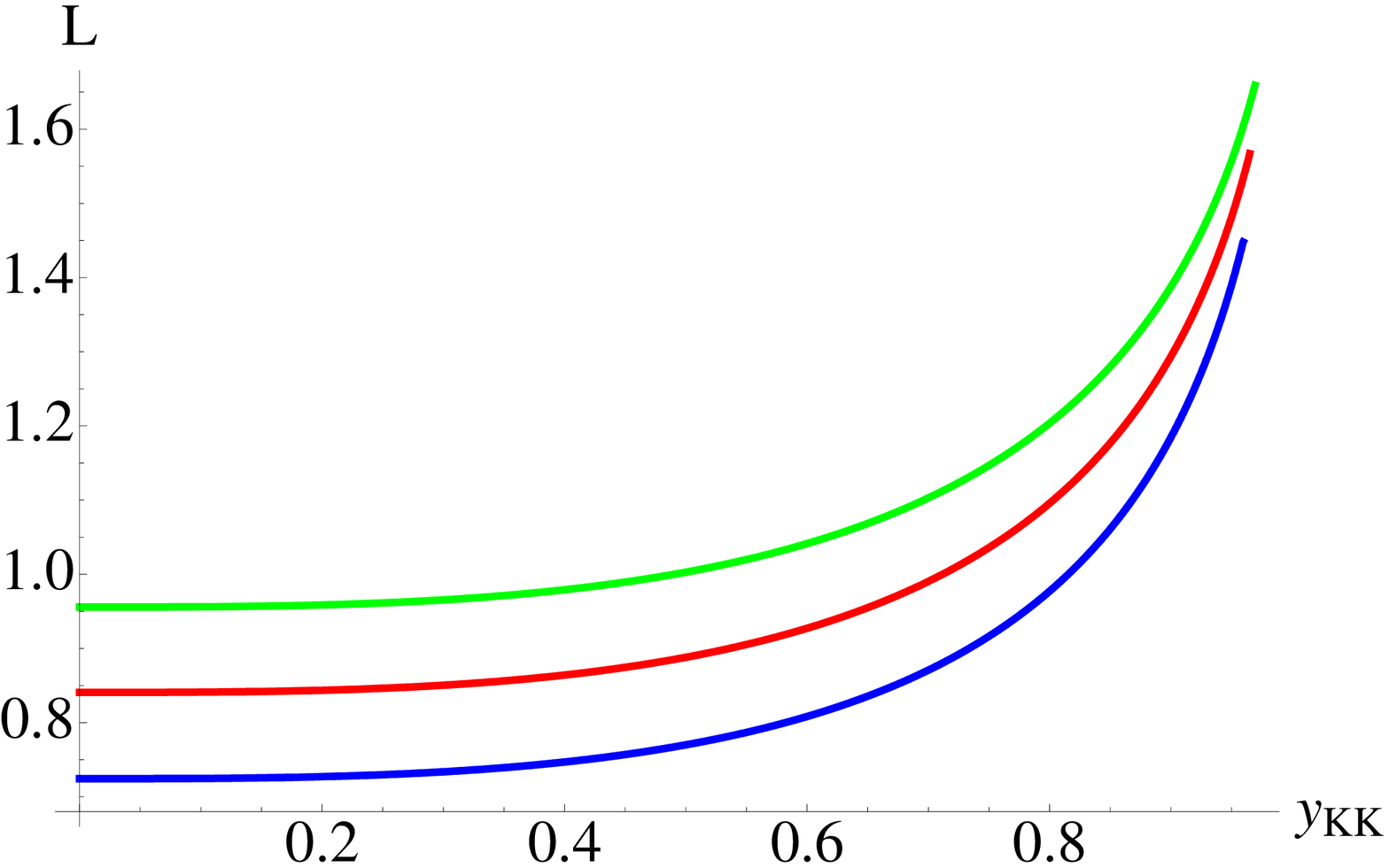} \label{fig: lengthlotyt}}
\subfigure[] {\includegraphics[angle=0,
width=0.45\textwidth]{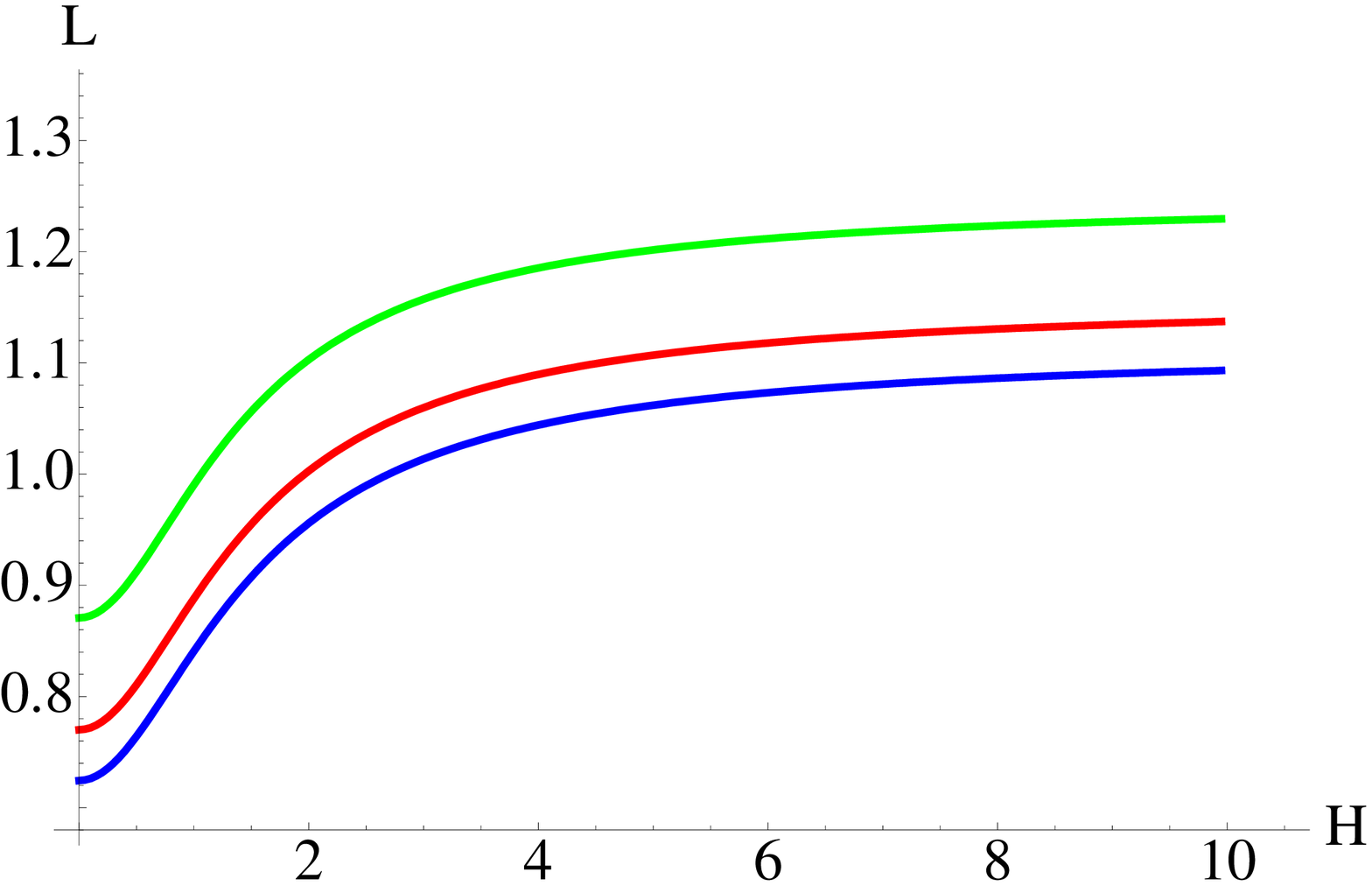} \label{fig: lengthlotmag}}
\caption{\small The dependence of the asymptotic separation between the flavour branes in units of $R_{D4}^{3/2}/\sqrt{U_0}$ with $y_{KK}$ and magnetic field in the low temperature phase. Figure \ref{fig: lengthlotyt} shows the dependence of $L$ with $y_{KK}$ for different magnetic field strength; blue (bottom most) curve corresponds to $H=0$, red (middle) corresponds to $H=1.0$, green (top most) corresponds to $H=2.0$. Figure \ref{fig: lengthlotmag} shows the behaviour of $L$~with applied magnetic field for different values of $y_{KK}$; blue (bottom most) curve corresponds to $y_{KK}=0$, red (middle) corresponds to $y_{KK}=0.5$, green (top most) corresponds to $y_{KK}=0.7$. We have set $d=1$.}
\end{center}
\end{figure}

Now small asymptotic separation corresponds to large values of $U_{0}$ which means $y_{KK}\ll 1$. So for small $L$ and weak magnetic field ($1/d^{3/2}\gg H$), the leading behaviour of the separation is given by (using equation~(\ref{eqt: sepz})), $L\sim R_{D4}^{3/2}/\sqrt{U_{0}}$. This is same as the leading behaviour in zero magnetic field case in ref.~\cite{Aharony:2006da}. However, for strong magnetic field ($1/d^{3/2}\ll H$), the leading behaviour obtained from equation~(\ref{eqt: sepz}) is given by, $L\sim R_{D4}^3 H/U_{0}^2$. So for fixed value of $U_{0}$ the asymptotic separation scales with the applied magnetic field strength $H$. This is however true only in the $y_{KK}\ll 1$ limit. The general dependence is more complicated and a numerical study yields figures \ref{fig: lengthlotyt} and \ref{fig: lengthlotmag}.

From figure \ref{fig: lengthlotyt} we can see that in the low temperature phase the asymptotic separation increases as the point of joining (i.e, $U_0$) of the flavour branes decreases. This means that the end points of the $D8/\overline{D8}$ move further and further away as we go deeper and deeper in the core. The role of magnetic field is to further increase this asymptotic distance for a given $y_{KK}$. However as we approach $y_{KK}=1$, the magnetic field does not affect the separation of the flavour branes any more, since all curves start converging rapidly near $y_{KK}=1$. This is the point where the background geometry ends precisely where the flavour branes join, therefore all probes for any asymptotic separation should end at this point irrespective of where they start from at infinity. This is consistent with earlier studies in ref.~\cite{Gepner:2006qy}. 

On the other hand, figure \ref{fig: lengthlotmag} shows that for a fixed value of $y_{KK}$ magnetic field can increase the asymptotic separation, but not indefinitely. This means that for a fixed $y_{KK}$ as we increase the magnetic field the flavour branes move further away from each other, but for high enough magnetic field this separation saturates and becomes insensitive to further increment of the magnetic field. We will see later that such saturation shows up in other physical quantities also. The role of $y_{KK}$ here is to shift each curve upwards as we increase its value. 

In ref.~\cite{Antonyan:2006vw} it was noted that the special case of $y_{KK}=0$ (meaning when the radius of the spatial circle goes to infinity and the compact direction becomes a flat direction) the supergravity background is dual to a non-local NJL model in which the separation scale between the brane--anti-brane pair (denoted as $L$ here) determines an effective coupling for a four fermi interaction term. As we have seen the magnetic field affects the asymptotic separation and therefore tunes the effective coupling.

Note that in this model the bare quark mass is always zero as there is no separation between the flavour and the colour branes at the boundary; however since the branes join at some length scale $U_0 \ge U_{KK}$ in the core one can consider a string stretching from $u=U_{KK}$ to $u=U_0$. The mass associated with the string can be identified to be the effective constituent quark mass as argued in ref.~\cite{Aharony:2006da}. If we denote this mass by $M_{q}$, then $M_{q}=\frac{1}{2\pi\alpha'}\int_{U_{KK}}^{U_0}\sqrt{g_{tt}g_{uu}}$\ , where $U_0$ has to be determined from equation~(\ref{eqt: sepz}) for given~$L$. This turns out to be a self-consistency equation for $U_0$. This is analogous to the Gap equation in the field theory context (e.g., in ref.~\cite{Antonyan:2006vw}). The self-consistency equation turns out to be
\begin{eqnarray}\label{eqt: lowgap}
&& U_0=\frac{4}{9}\frac{R_{D4}^3}{L^2}\left(1-y_{KK}^3\right)\left(U_0^3+H^2R_{D4}^3\right) I(U_0,H)^2\ ,\quad {\rm where} \quad\nonumber\\
&& I(U_0,H)=\int_0^1\frac{\left(1-y_{KK}^3z\right)^{-1}z^{-5/6}dz}{\left(z^{-8/3}\left(1-y_{KK}^3z\right)\left(U_0^3+H^2R_{D4}^3\right)-\left(1-y_{KK}^3\right)\left(U_0^3+H^2R_{D4}^3\right)\right)^{1/2}}\ .\nonumber\\
\end{eqnarray}

Now equation~(\ref{eqt: lowgap}) can be solved perturbatively, i.e.,
starting with an initial value for the parameter $U_0$ we can
determine the next order approximation to $U_0$ using
equation~(\ref{eqt: lowgap}); and we continue until the desired
accuracy has been achieved. It is straightforward to guess the initial
value of $U_0$. Plugging in $H=0$ in equation~(\ref{eqt: lowgap}) we
should get the constituent mass for the low temperature case. This can
serve as the initial guess for small magnetic fields. Once $U_0$ is
known for small magnetic fields, it can be used as the initial guess
for successively higher values of magnetic fields. Thus we obtain the
dependence which is shown in figure below.

\begin{figure}[!ht]
\begin{center}
\includegraphics[angle=0,
width=0.55\textwidth]{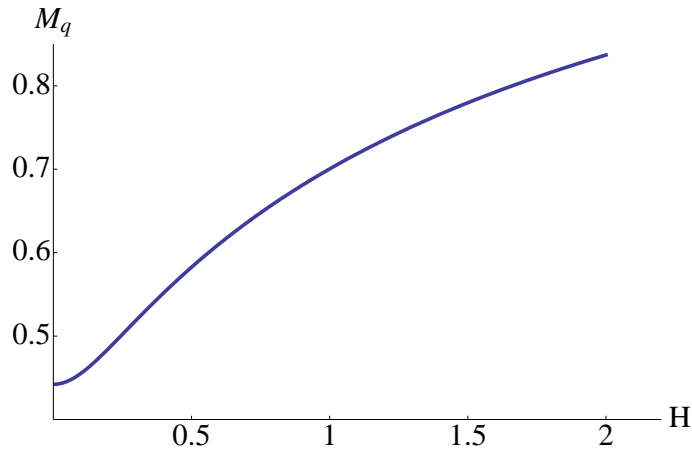}
\caption{\small The dependence of constituent quark mass (measured in units of $(2\pi\alpha')^{-1}$) on the external magnetic field in the low temperature phase. We have fixed $U_{KK}=0.4$ and $R_{D4}=1$.}
\label{fig: mdyn0}
\end{center}
\end{figure}

It is expected that the mass of the vector and pseudoscalar meson
would monotonically increase as the constituent quark mass increases.
Figure \ref{fig: mdyn0} therefore should capture the behaviour of
massive vector and pseudoscalar meson spectra in presence of magnetic
field.
  
Note that for this background disjoint brane pair do not exist: The
constant $\tau$-solutions namely, $\tau=-\pi R/2$ and $\tau=+\pi R/2$
join at $u=U_{KK}$. This is because of the cigar geometry of the
background in the $\{\tau,u\}$ submanifold and the fact that $\tau$-circle
is wrapped by the probe branes. The brane pair must join together
since there is no place in the geometry for them to end separately. So
the only configuration possible in the low temperature phase breaks
chiral symmetry by reducing the global $U(N_f)_L\times U(N_f)_R$ to the
diagonal $U(N_f)$. This is geometrically understood as the joining of
the flavour $8$-branes.

We can compute the action for this brane by substituting for
the profile function into equation~(\ref{eqt: dbi}). The result is
\begin{eqnarray}\label{eqt: energyz}
S_{D8} 
     = \frac{2C}{3}R_{D4}^{3/2}U_{0(H)}^{7/2}\int_0^1 \frac{z^{-8/3}\left(1+zH^2d^3\right)}{\left[\left(1-y_{KK}^3z\right)\left(1+zH^2d^3\right)-z^{8/3}\left(1-y_{KK}^3\right)\left(1+H^2d^3\right)\right]^{1/2}}\ .\nonumber\\
\end{eqnarray}

Note that for small $L$, this energy is proportional to the vacuum
energy of the system. Recalling the behaviour of $L$ for small $y_{KK}$
and weak magnetic field we get $S_{D8}\sim 2CR_{D4}^{3/2}/L^7$. This
is similar to the leading order behaviour in absence of magnetic field
obtained in ref.~\cite{Aharony:2006da}. On the other hand, for small
$y_{KK}$ and strong magnetic field we get $S_{D8}\sim
2CR_{D4}^3H^{1/2}/L^{7/2}$. So sufficiently high magnetic field
changes how the vacuum energy blows up as $L\to 0$. The general
dependence is more involved which we do not pursue here.

\subsection{The High Temperature Background}

\begin{figure}[!ht]
\begin{center}
\includegraphics[angle=0,
width=0.65\textwidth]{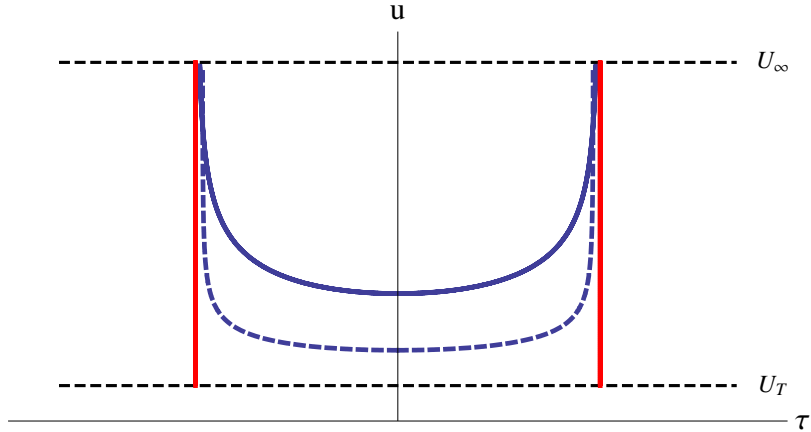}
\caption{\small The dashed U-shaped curve represents a profile in vanishing background field and the solid U-shaped curve represents a profile when a non-zero magnetic field is present. The straight (red) solution does not have any qualitative change in presence or absence of the external field. These profiles are obtained by numerically solving the equation of motion for the probe brane.}
\label{fig: phigh}
\end{center}
\end{figure}

Recall that the high temperature background is given by
equation~(\ref{eqt: highmet}); and we again use the same ansatz for the probe given
by equation~(\ref{eqt: ansatz}). As before the profile of the probe brane is completely determined by the DBI action, from which a first integral of motion can be readily obtained to be
\begin{eqnarray}\label{eqt: highfirstint}
u^4\frac{\left(1+H^2\left(\frac{R_{D4}}{u}\right)^3\right)^{\frac{1}{2}}f(u)}{\left(f(u)+\left(\frac{R_{D4}}{u}\right)^3u'^2\right)^{\frac{1}{2}}}=U_0^4\left(1+H^2\left(\frac{R_{D4}}{U_0}\right)^3\right)^{\frac{1}{2}}\sqrt{f(U_0)}\ .
\end{eqnarray}
For convenience we use the dimensionless variables defined in equation~(\ref{eqt: rescale}) along with the new variable $y_T=U_T/U_0$. Starting from the first integral of motion in equation~(\ref{eqt: highfirstint}) it is easy to verify that the finite temperature analogue to equation~(\ref{eqt: diffslope}) takes the following form
\begin{eqnarray}\label{eqt: newcom}
u_H'^2-u_{H=0}'^2=\left(\frac{y}{d}\right)^3\frac{f(y)^2}{f(1)}y^8\frac{H^2d^3}{1+H^2d^3}\left(\frac{1}{y^3}-1\right)\ .
\end{eqnarray}
This also suggests that $|u_H'|\le |u_{H=0}'|$, leading us to the same conclusion that the magnetic field helps bending the branes. The analogue to equation~(\ref{eqt: sepmag}) and~(\ref{eqt: sepnomag}) now take the following forms
\begin{eqnarray}\label{eqt: newlen}
&& \frac{L}{2}=\frac{R^{3/2}}{\sqrt{U_{0(H)}}}\int_1^\infty \frac{y^{-3/2}dy}{\sqrt{f(y)}\left[\frac{f(y)}{f(1)}\frac{1+H^2\left(\frac{d}{y}\right)^3}{1+H^2d^3}y^8-1\right]^{1/2}}=\int_1^\infty \mathcal{I}_H dy\ ,\nonumber\\
&& \frac{L}{2}=\frac{R^{3/2}}{\sqrt{U_{0(H)}}}\int_1^\infty \frac{y^{-3/2}dy}{\sqrt{f(y)}\left[\frac{f(y)}{f(1)}y^8-1\right]^{1/2}}=\int_1^\infty \mathcal{I}_0 dy\ ,
\end{eqnarray}
leading us to a similar conclusion as the low temperature case. This is pictorially represented in figure \ref{fig: phigh}.

\begin{figure}[h!]
\begin{center}
\subfigure[] {\includegraphics[angle=0,
width=0.45\textwidth]{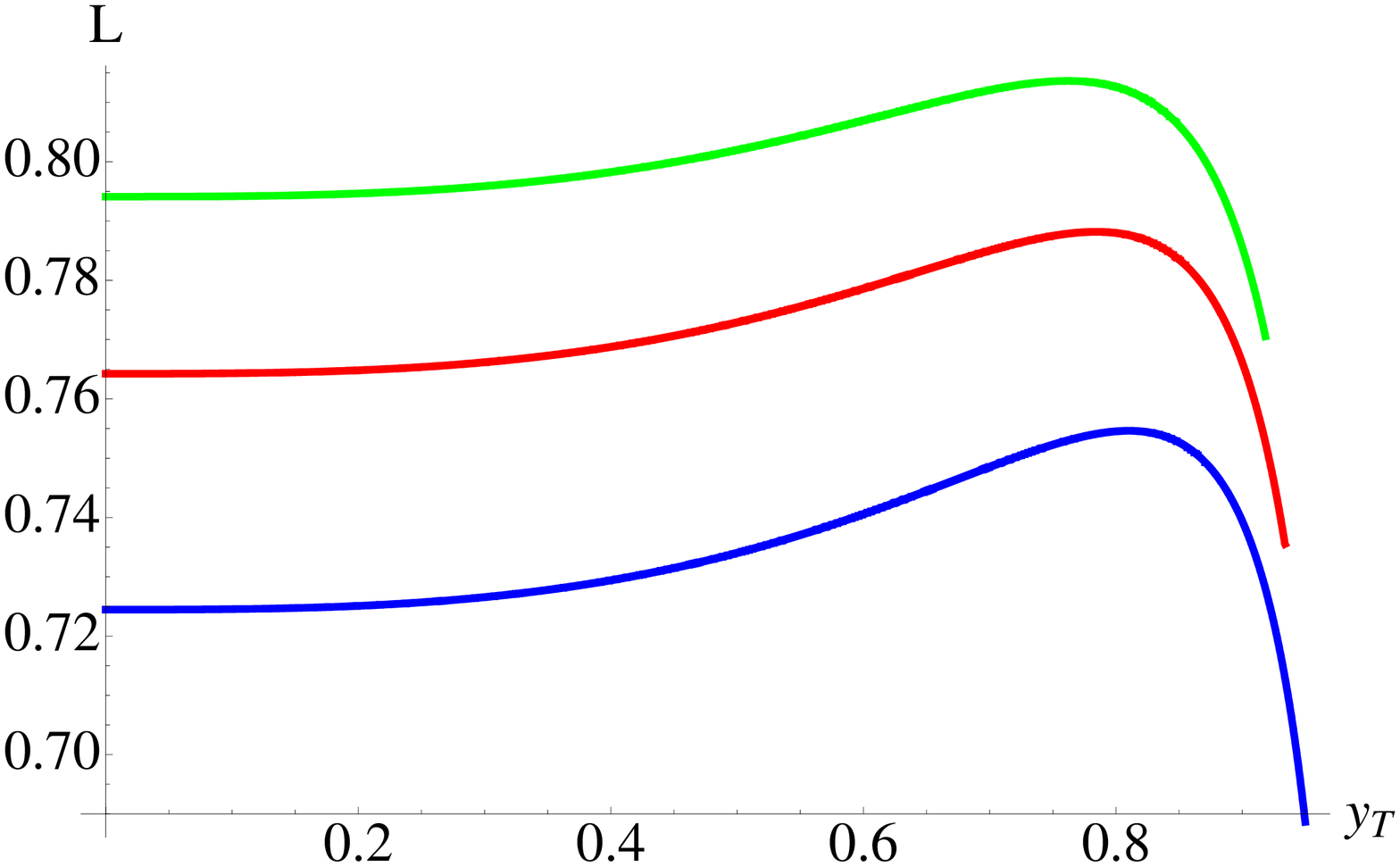} \label{fig: lengthhightyt}}
\subfigure[] {\includegraphics[angle=0,
width=0.45\textwidth]{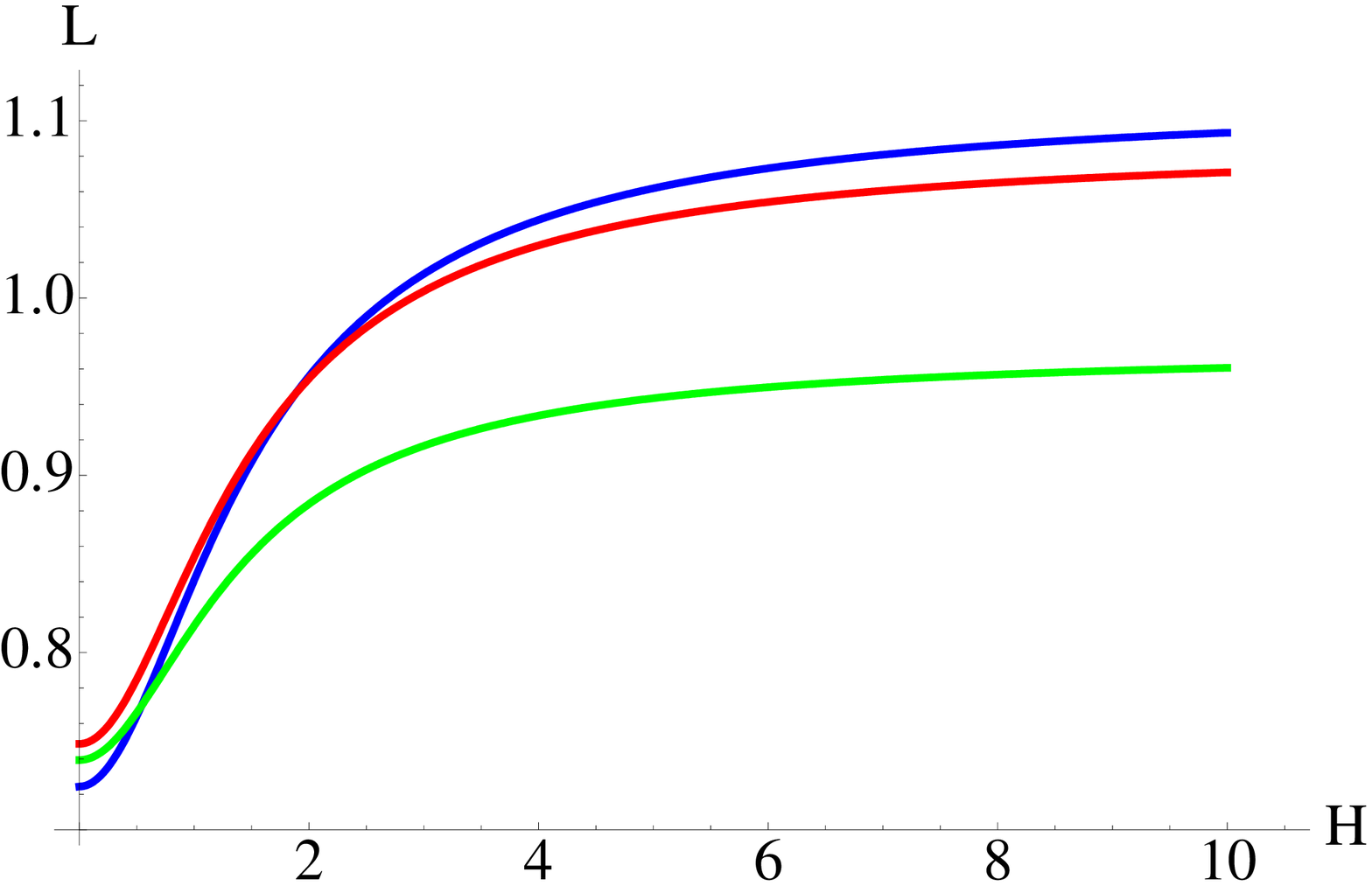} \label{fig: lengthhightmag}}
\caption{\small The dependence of asymptotic separation between the flavour branes in units of $R_{D4}^{3/2}/\sqrt{U_0}$ with $y_T$ and magnetic field in the high temperature limit. Figure \ref{fig: lengthhightyt} shows the dependence of $L$ with $y_T$ for different magnetic field strength; blue (bottom most) curve corresponds to $H=0$, red (middle) corresponds to $H=0.5$, green (top most) corresponds to $H=0.7$. Figure \ref{fig: lengthhightmag} shows the behaviour of $L$ with applied magnetic field for different values of $y_T$; blue (top most from right) curve corresponds to $y_T=0$, red (middle from right) corresponds to $y_T=0.7$, green (bottom most from right) corresponds to $y_T=0.9$. We have set $d=1$.}
\end{center}
\end{figure}

As before one can extract the dependence of the asymptotic separation in the small $y_T$ limit. Again with the change to variable $z=y^{-3}$ we get
\begin{eqnarray}\label{eqt: sepzmag}
\frac{L}{2}=\frac{R_{D4}^{3/2}}{3\sqrt{U_{0(H)}}}\sqrt{\left(1-y_T^3\right)\left(1+H^2d^3\right)}\int_0^1 \frac{\left(1-y_T^3z\right)^{-1/2} z^{-5/6}dz}{\sqrt{\frac{\left(1-y_T^3z\right)\left(1+H^2d^3z\right)}{z^{8/3}}-\left(1-y_T^3\right)\left(1+H^2d^3\right)}}\ .\\ \nonumber
\end{eqnarray}
So from equation~(\ref{eqt: sepzmag}) one can see that for small $y_T$ and weak magnetic field $L\sim R_{D4}^{3/2}/\sqrt{U_{0(H)}}$; whereas for small $y_T$ and strong magnetic field we get $L\sim R_{D4}^3H/U_{0(H)}^2$. The general dependence has been studied numerically and the result is shown in figure \ref{fig: lengthhightyt} and \ref{fig: lengthhightmag}.

In figure \ref{fig: lengthhightyt} we see that the asymptotic
separation decreases as $y_T$ increases, consistent with studies in
ref.~\cite{Gepner:2006qy}. The role of magnetic field is to shift the
curves upwards, namely to increase the asymptotic separation. However
this effect vanishes as $y_T$ approaches its maximum value and the
separation becomes insensitive to the background magnetic field.
Figure \ref{fig: lengthhightmag} shows a similar behaviour as the low
temperature case. The separation at the boundary becomes higher and
higher for increasing magnetic field, but for high enough field the
flavour branes at the boundary tend to not sense any further increment
(therefore a saturation is obtained). In this case however, curves for
different $y_T$ may intersect each other (as shown in figure \ref{fig:
  lengthhightmag}) unlike the low temperature case.
 
\begin{figure}[!ht]
\begin{center}
\includegraphics[angle=0,
width=0.55\textwidth]{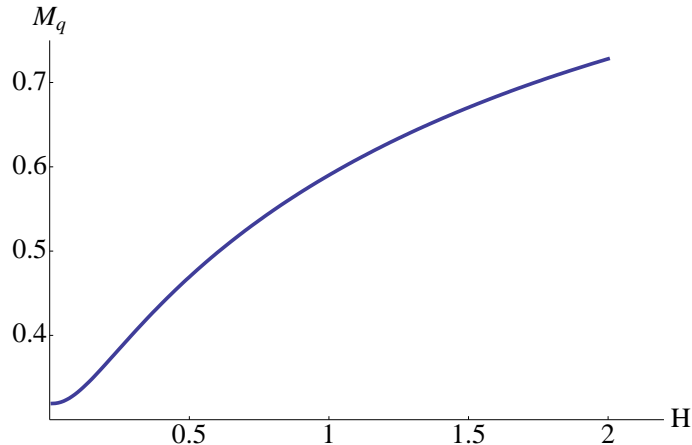}
\caption{\small The dependence of constituent quark mass (measured in units of $(2\pi\alpha')^{-1}$) on the external magnetic field in the high temperature phase. We have fixed $U_{T}=0.3$ and $R_{D4}=1$.}
\label{fig: mdynT}
\end{center}
\end{figure}

The joining of the flavour branes inside the core can be associated
with the effective constituent quark mass. This corresponds to a
self-consistency equation for $U_0$ as in the low temperature case.
The equation can again be solved using the same perturbative approach
and the results are summarised in the figure \ref{fig: mdynT}. This
behaviour of constituent mass is valid when the curved solutions are
the lowest energy solutions (the chiral symmetry broken phase). In the
presence of finite temperature there will be a first order transition
to chiral symmetry restored phase. We will study this transition later
in the next section.

Now the trivial solution of equation~(\ref{eqt: highfirstint}), which
is given by $\tau'=0$ has a different physical meaning from the point
of chiral symmetry breaking. Since the $\{t,u\}$ submanifold has the
cigar geometry, the solutions $\tau=\pm L/2$ can remain disjoint and
end at $u=U_T$ separately. Therefore the trivial solutions in the high
temperature case preserve the full $U(N_f)_L\times U(N_f)_R$ symmetry
by remaining disjoint. In order to determine the true minimum energy
configuration we need to compare the energies of the curved and the
straight branes. We pursue this in the next section.

\subsubsection{The Probe Brane Profile and Chiral Symmetry Breaking}
 
 To determine the true vacuum
we consider the difference between the energies of the curved and straight branes, which is given by
\begin{eqnarray}\label{eqt: dissmag}
\Delta S =\frac{S_{\rm curved}-S_{\rm straight}}{C U_0^5d^{\frac{3}{2}}}
         &=& \int_1^\infty dy y\left(y^3+H^2d^3\right)^{1/2}\left[\frac{1}{\left(1-\frac{f(1)(1+H^2d^3)}{f(y)(y^3+H^2d^3)}y^{-5}\right)^{1/2}}-1\right]\nonumber\\
       && -\int_{y_T}^1 dy y\left(y^3+H^2d^3\right)^{1/2}\ .
\end{eqnarray}
Here $\Delta S<0$ would mean chiral symmetry breaking, $\Delta S>0$ would mean chiral symmetry restoration and $\Delta S=0$ would characterize a transition from symmetry broken phase to a symmetry restored one.
We employ numerical analysis to study this. 

\begin{figure}[h!]
\begin{center}
\subfigure[] {\includegraphics[angle=0,
width=0.45\textwidth]{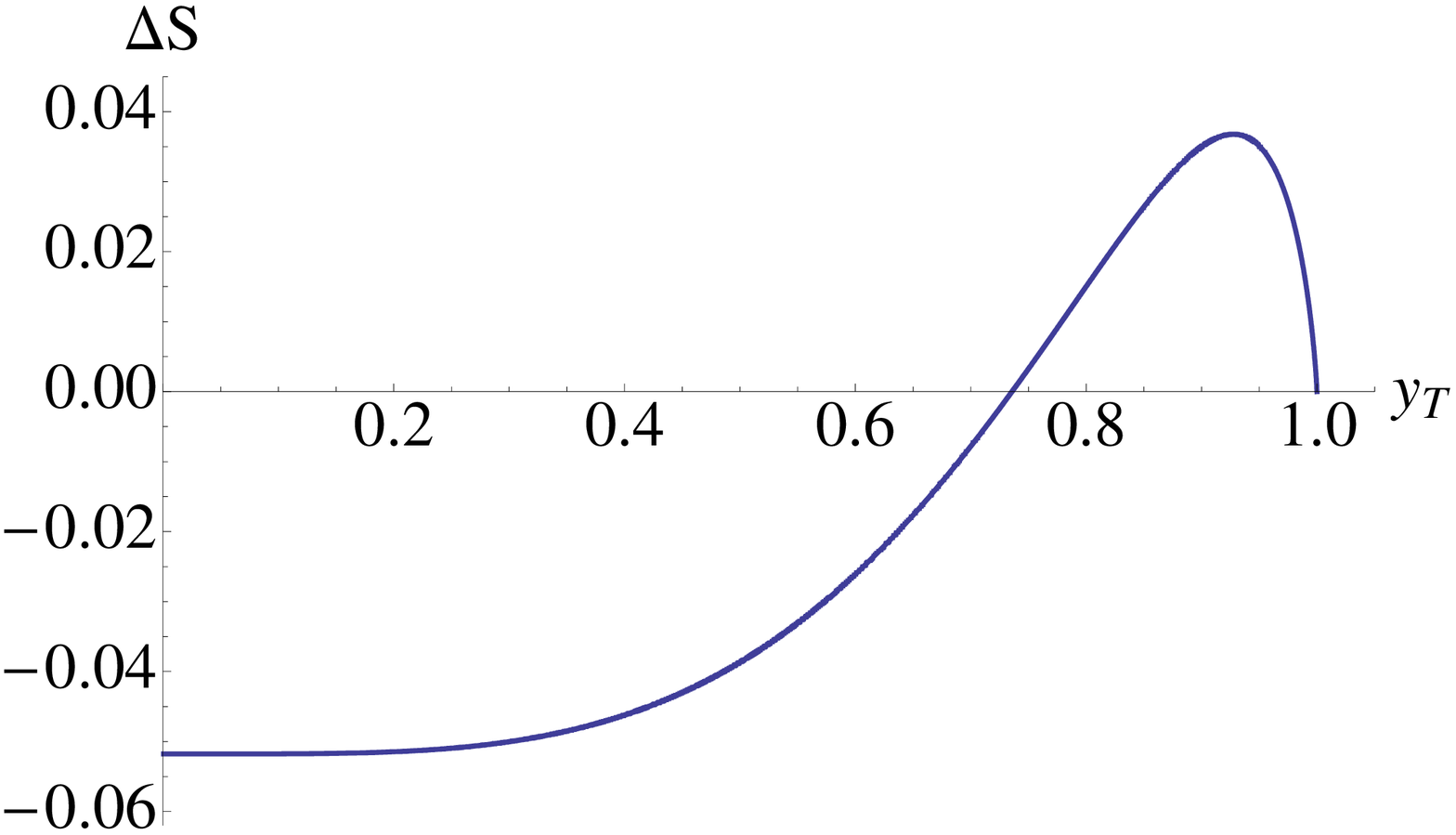} \label{fig: difft}}
\subfigure[] {\includegraphics[angle=0,
width=0.45\textwidth]{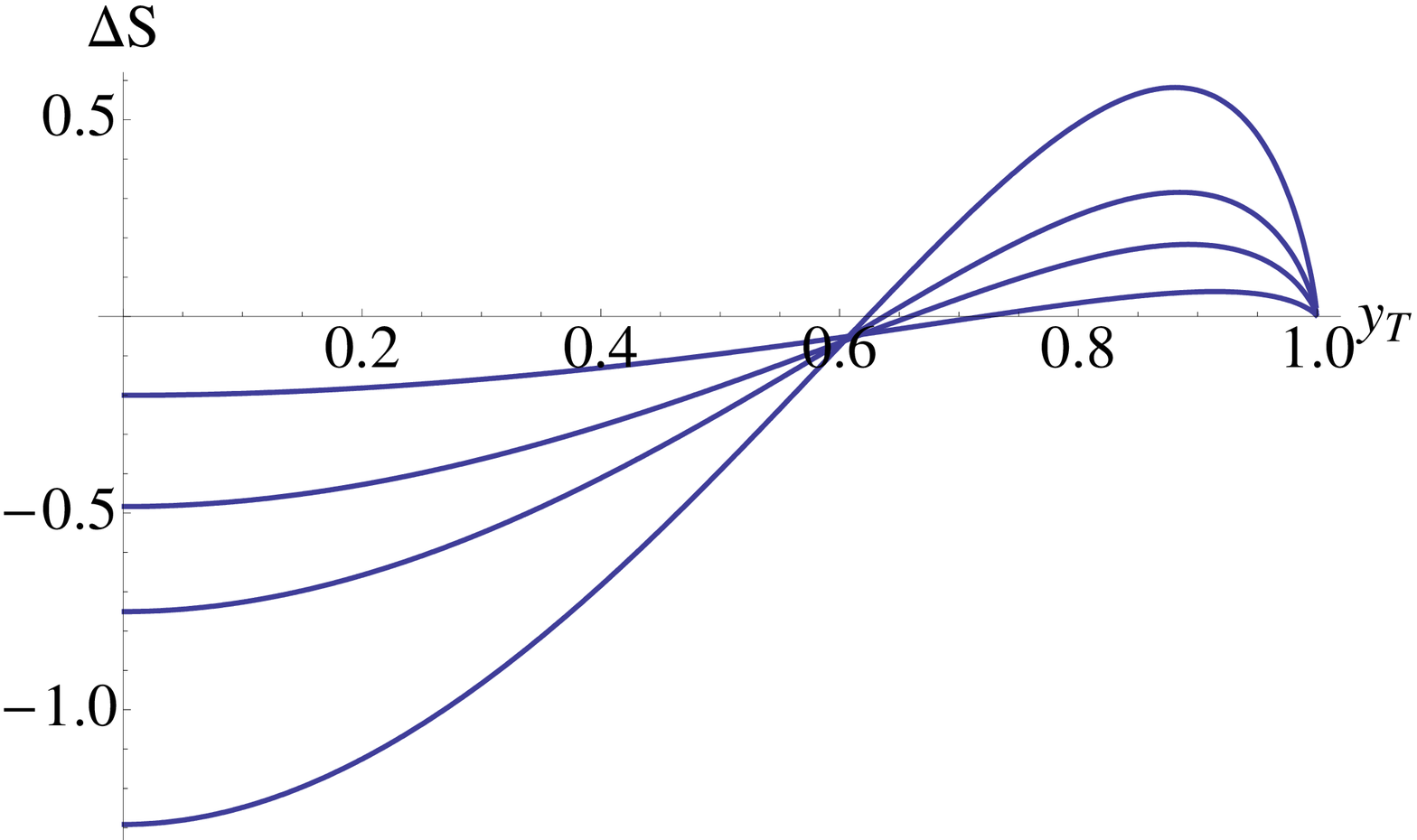} \label{fig: diffmag}}
\caption{\small The dependence of zeros of $\Delta S$ on the magnetic field. Figure \ref{fig: difft} shows the result for zero magnetic field and figure \ref{fig: diffmag} shows how the zero changes as we fix $H=1,\, 3,\, 5,\, 9$ from above to below respectively. We have set $d=1$.}
\end{center}
\end{figure}

It is known from, e.g.,
ref.~\cite{Aharony:2006da} that for high temperature and zero magnetic
field there exists a critical temperature beyond which the straight
branes are energetically favoured, implying that in the dual gauge
theory chiral symmetry is restored. Below the temperature, however,
the symmetry is broken by energetically favoured curved brane pair
that join together. The plots for the energy difference $\Delta S$ is
shown in figures \ref{fig: difft} and \ref{fig: diffmag} for zero and
non--zero values of magnetic field respectively.
 
\begin{figure}[!ht]
\begin{center}
\includegraphics[angle=0,
width=0.55\textwidth]{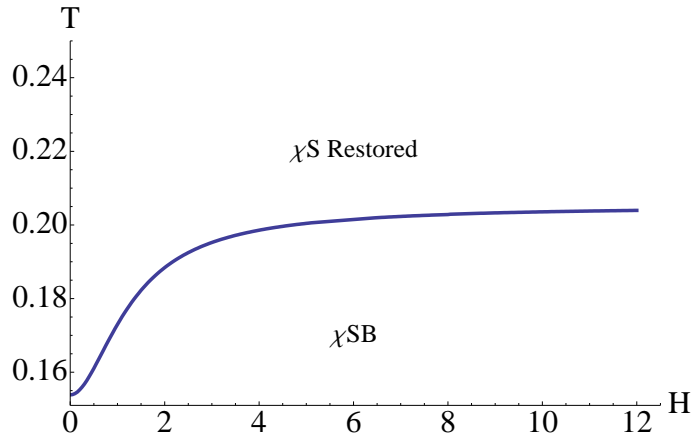}
\caption{\small The phase diagram between applied magnetic field and the chiral symmetry restoring temperature $T_{\chi SB}$ in units of $1/L$, where $L$ is the asymptotic separation between the branes. We have set $d=1$.}
\label{fig: phase}
\end{center}
\end{figure}

It is evident that the first order phase transition from chiral
symmetry broken to the symmetry restored phase persists in presence of
external magnetic field. From the zero of $\Delta S$ we can find out
the critical value of $y_T$ for which the symmetry restoration occurs.
Now to represent the phase diagram in terms of physical quantities, we
recall that there is a length scale $L$ corresponding to the
separation of the brane--anti-brane pair at the boundary. So we
express the chiral symmetry restoring temperature in units of $1/L$
using the critical value of $y_T$ in equation~(\ref{eqt: sepzmag}). The
resulting phase diagram is shown in figure~\ref{fig: phase}.

It is interesting to note that the presence of magnetic field
increases the symmetry restoring temperature. In other words it
promoted the spontaneous breaking of chiral symmetry. This fits with
the general expectations from field theory (see e.g. refs.\cite{Miransky:2002eb}) and
the supergravity/probe intuition that introducing a magnetic field
places more energy into the system; therefore in order to minimize the
energy, condensates are formed (the branes bend more) resulting in
more readily broken chiral symmetry. (It should be noted, however,
that in this specific holographic model the identification of a quark
condensate is a rather subtle issue (see e.g.,
ref.~\cite{Bergman:2007pm}).)

We can extract some more information about the transition by studying
certain thermodynamic quantities at the phase transition. To that end,
let us note that the first order phase transition is accompanied by
entropy density that jumps at $T=T_c$ yielding a non--zero latent heat
as reported in ref.~\cite{Parnachev:2006dn}, also a change in
magnetization
\begin{eqnarray}
&& \Delta s=-\frac{1}{V_{R^3}}\frac{\partial\left(S_{\rm curved}-S_{\rm straight}\right)}{\partial T}\ , \quad C_{latent}=T_c\Delta s\ ,\nonumber\\
&& \Delta\mu=-\frac{1}{V_{R^3}}\frac{\partial\left(S_{\rm curved}-S_{\rm straight}\right)}{\partial H}\ .
\end{eqnarray}

\begin{figure}[h!]
\begin{center}
\subfigure[] {\includegraphics[angle=0,
width=0.45\textwidth]{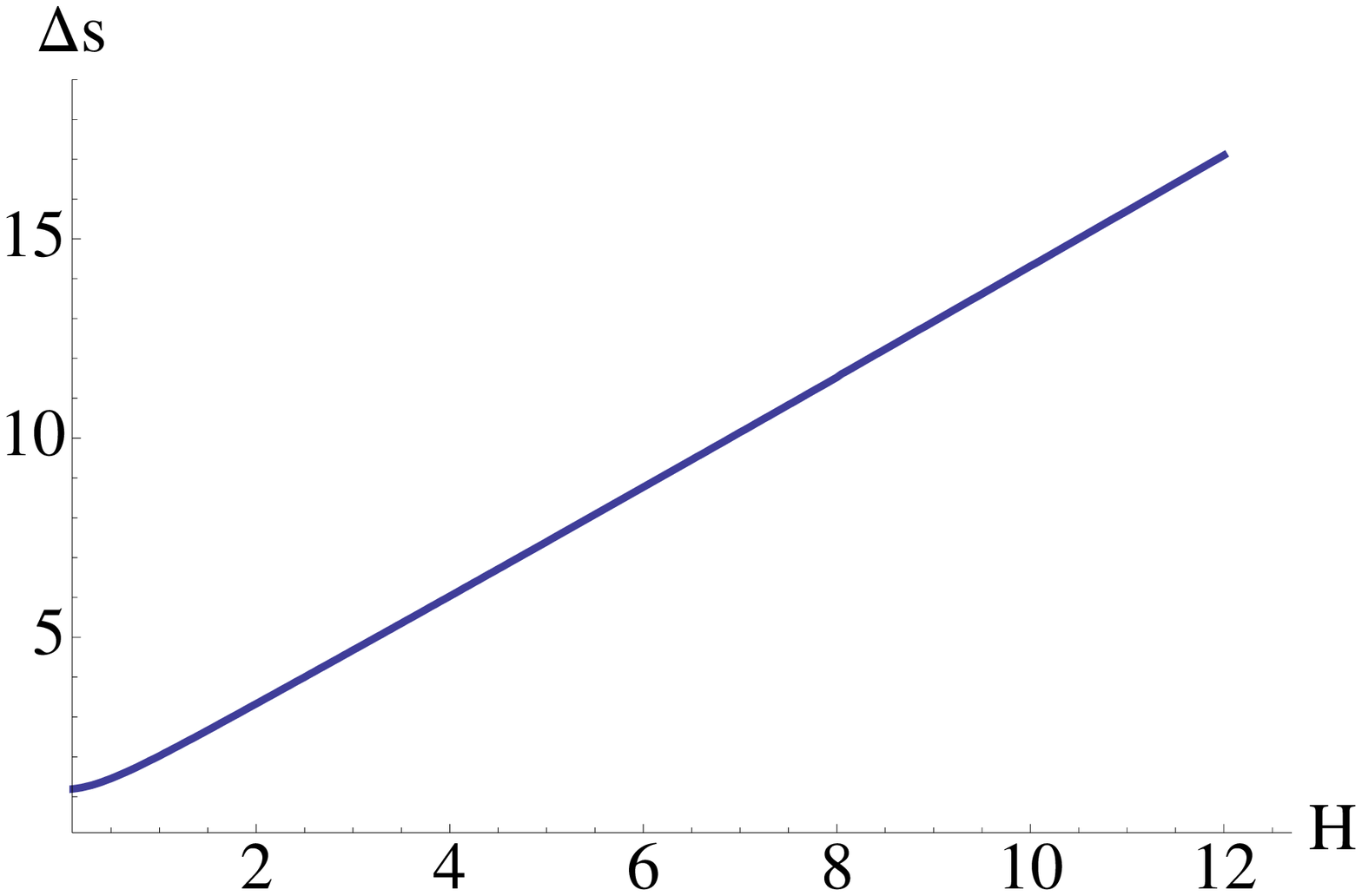} \label{fig: scrit}}
\subfigure[] {\includegraphics[angle=0,
width=0.45\textwidth]{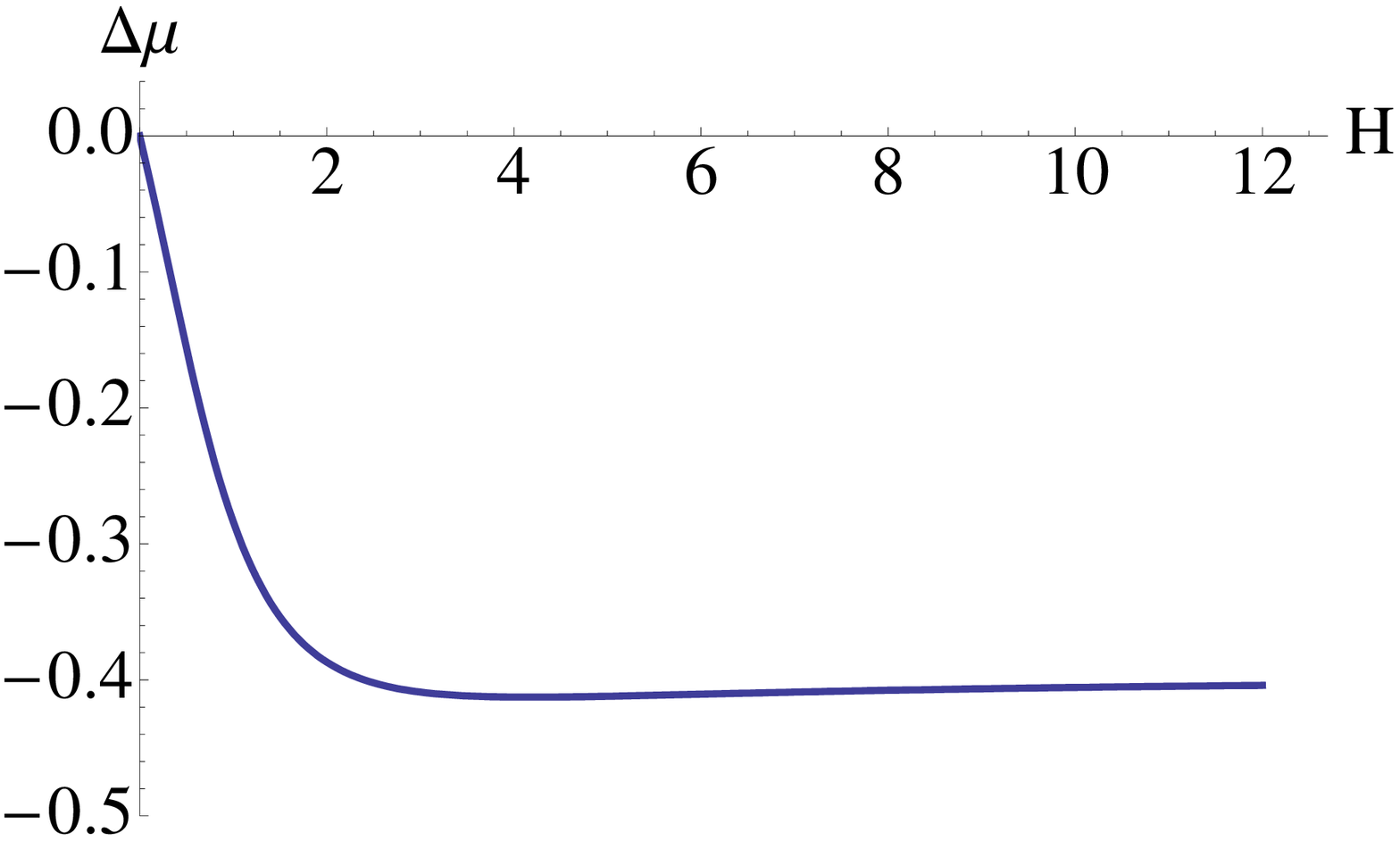} \label{fig: mucrit}}
\caption{\small The behaviour of jump in entropy and change in magnetisation at the critical temperature where chiral symmetry is being restored. The vertical axis is evaluated in units of $CU_0^{7/2}R_{D4}^{3/2}$ and we have set $d=1$.}
\end{center}
\end{figure}

The absolute free energy and any thermodynamic quantity obtained from it (such as the absolute magnetization) for the two classes of embeddings (the straight and the curved branes respectively) are formally divergent quantities. Hence we compute the relative quantities which are finite. We studied the dependence of the change in entropy density and the relative magnetization numerically, and the
results are shown in figures \ref{fig: scrit} and \ref{fig: mucrit}
respectively. The relative magnetization also shows a similar
saturation behaviour for high enough magnetic field. The straight branes correspond to the melted phase where quarks are free whereas the curved branes correspond to the mesonic phase where quarks exists in the form of bound states or chiral condensates. Therefore it is expected that the chiral symmetry restored phase (corresponding to the straight branes) is more ionized than the chiral symmetry broken phase (corresponding to the curved branes). This is in accord with our observation that the relative magnetization is negative in figure \ref{fig: mucrit}.

\section{Probing with $Dp$-Brane}

We now consider general $Dp$/$\overline{Dp}$-brane as flavours in the
colour $D4$-brane background. The low temperature background is given
by equation~(\ref{eqt: metsakai}) and the high temperature is given by
equation~(\ref{eqt: highmet}). Also we choose to parametrize the $S^4$
as follows
\begin{eqnarray}
 d\Omega_4^2=d\theta_1^2+\sin^2\theta_1d\theta_2^2+\sin^2\theta_1\sin^2\theta_2\left(d\theta_3^2+\sin^2\theta_3d\theta_4^2\right)\ .
\end{eqnarray}
There are several ways to place the probe brane. We list these
possibilities in the table below (being somewhat cavalier with the use of the term QCD), as done in ref.~\cite{Gepner:2006qy}

\begin{center}\label{table: poss}
\begin{tabular}{|r|r|r|r|r|r|r|r|r|r|r|r|}
\hline
 &$t$&$x^1$&$x^2$&$x^3$&$\tau$&$u$&$\theta_1$&$\theta_2$&$\theta_3$&$\theta_4$& \\
\hline
$D4$&$-$&$-$&$-$&$-$&$-$&$\times$&$\times$&$\times$&$\times$&$\times$& \\
\hline
\hline
$D4$&$-$&$-$&$-$&$-$&$\times$&$-$&$\times$&$\times$&$\times$&$\times$&$QCD_4$\\
$D6$&$-$&$-$&$-$&$-$&$\times$&$-$&$\times$&$\times$&$-$&$-$&$QCD_4$\\
$D8$&$-$&$-$&$-$&$-$&$\times$&$-$&$-$&$-$&$-$&$-$&$QCD_4$\\
\hline
\hline
$D4$&$-$&$-$&$-$&$\times$&$\times$&$-$&$\times$&$\times$&$\times$&$-$&$QCD_3$\\
$D6$&$-$&$-$&$-$&$\times$&$\times$&$-$&$\times$&$-$&$-$&$-$&$QCD_3$\\
\hline
\hline
$D2$&$-$&$-$&$\times$&$\times$&$\times$&$-$&$\times$&$\times$&$\times$&$\times$&$QCD_2$\\
$D4$&$-$&$-$&$\times$&$\times$&$\times$&$-$&$\times$&$\times$&$-$&$-$&$QCD_2$\\
$D6$&$-$&$-$&$\times$&$\times$&$\times$&$-$&$-$&$-$&$-$&$-$&$QCD_2$\\
\hline
\end{tabular}
\end{center}

To introduce flavour brane in this set up we need to introduce probe
branes that extend in the $u$ direction all the way up to $u=\infty$.
In general we consider introducing $Dp-\overline{Dp}$ brane which
intersect the colour $D4$--branes at $\tau=\pm L/2$ at the boundary,
where the $\mp$ sign corresponds to position of the $\overline{Dp}$ and
$Dp$-branes respectively. It can be noted that the coordinate $\tau$
ranges from $-\pi R$ to $+\pi R$, and therefore $L\in [-\pi R/2, +\pi
R/2]$. By construction the flavour symmetry group\footnote{For the $(2+1)$-dimensional case, chiral fermions do not exist. Therefore the global symmetry is just an $U(N_f)\times
U(N_f)$ which we continue to refer as the chiral symmetry.} is $U(N_f)_L\times
U(N_f)_R$.  Depending on the dimension of the probe brane, the number
of common directions between the colour and the flavour branes is
determined, which corresponds to the dimension of the dual gauge
theory. Such construction has previously been discussed in
ref.~\cite{Gepner:2006qy}. We follow the same approach as before to
introduce magnetic field. By construction we can see that the magnetic
field which has support along the $(x^2,x^3)$ direction will have no
effect on $QCD_2$.

\subsection{Low Temperature Phase}

We use similar ansatz as in equation~(\ref{eqt: ansatz}) for
$Dp$/$\overline{Dp}$-brane, however depending on the dimension of the
probe brane we now place it at $\theta_i=\pi/2$ as required following
the table. We also denote the profile of the probe brane by $\tau(u)$
(we change the choice of function representing the brane profile,
earlier we considered to parametrize them as $u(\tau)$). The probe
brane DBI action\footnote{Once again it is straightforward to verify that there will be no Wess-Zumino term in any of these cases.} can be written in the following generic manner
\begin{eqnarray}\label{eqt: gendbi}
S_{Dp}=\frac{\mathcal{N}}{T}\int du e^{-\phi}\left(g_{tt}g_{uu}g_{xx}^{n-3}(g_{xx}^2+H^2)(det S_{p-n})\right)^{1/2}\ ,
\end{eqnarray}
where $g_{ab}$ are the induced metric components on the world-volume
of probe brane; $n$ is the dimension of the gauge theory and $(p+1)$
is the dimension of the probe $Dp$-brane. $T$ is the temperature of
the background and $\mathcal{N}$ is given by
\begin{eqnarray}
\mathcal{N}=\mu_p V_{R^n}V_{S^{p-n}}\ .
\end{eqnarray}
Similar to the $D8$-brane case it is trivial to see that this action
is independent of $\tau(u)$ and contains only the first derivative of
$\tau(u)$ with respect to $u$. Therefore the corresponding first
integral of motion will be the momentum corresponding to $\tau(u)$,
which is:
\begin{eqnarray}\label{eqt: genfirst}
\frac{u^{p-n}\left(\frac{u}{R_{D4}}\right)^{\frac{3}{4}(2n-p)}\left(1+H^2\left(\frac{R_{D4}}{u}\right)^3\right)^{\frac{1}{2}}f(u)\tau'}{\left(\tau'^2f(u)+\left(\frac{u}{R_{D4}}\right)^3\frac{1}{f(u)}\right)^{1/2}}={\rm const.} \nonumber
\end{eqnarray}
We see that this equation has the general structure as the $D8$-brane
considered in equation~(\ref{eqt: firstint}); only difference appears
in the power of an overall factor of $u$, which now contains the
general information of the dimension of the probe brane and the gauge
theory. Therefore magnetic field would have similar effects on the
profile by bending them more, promoting spontaneous  chiral symmetry breaking.

The action can be computed by substituting $\tau'$ from
equation~(\ref{eqt: genfirst}) into equation~(\ref{eqt: gendbi}). The
result is
\begin{equation}
S_{Dp}=\frac{\mathcal{N}d^{\frac{3}{2}}U_0^{p+1-n}}{g_sTd^{\frac{3}{4}(2n-p)}}\int_1^\infty dy \frac{y^{\frac{p+2n}{2}}\left(1+H^2\left(d/y\right)^3\right)y^{-3/2}}{\left(y^{\frac{p+2n}{2}}f(y)\left(1+H^2\left(d/y\right)^3\right)-f(1)(1+H^2d^3)\right)^{1/2}}\ ,\nonumber\\
\end{equation}
where we have used the rescaled variables as denoted in equation~(\ref{eqt: rescale}). The asymptotic separation of the brane--anti-brane system is again given by
${L/2}=\int_1^\infty dy \tau'\ .$

The physics of chiral symmetry breaking remains the same. As before, in the confined phase chiral symmetry is always broken which is enforced by the cigar geometry in $\{\tau,u\}$ submanifold of the background geometry. Next we consider the high temperature case and study the effect of magnetic field for a general probe brane.

\subsection{High Temperature Phase}

We carry out exactly similar analysis as before. We now find two distinct class of solutions, namely the straight branes and the curved branes. The energy of the curved branes is given by
\begin{eqnarray}\label{eqt: dpu}
S_{Dp}^{\rm curved}=\frac{2\mathcal{N}U_0^{p+1-n}d^{3/2}}{g_sTd^{\frac{3}{4}(2n-p)}}\int_0^1 dz \frac{z^{-\frac{4}{3}}z^{-\frac{p+2n}{6}}\left(1+H^2d^3z\right)z^{\frac{1}{2}}\sqrt{(1-y_T^3z)}}{\left(z^{-\frac{p+2n}{6}}\left(1+H^2d^3z\right)(1-y_T^3z)-(1+H^2d^3)(1-y_T^3)\right)^{1/2}}\ . \nonumber\\
\end{eqnarray}
The the energy of the straight branes is given by
\begin{eqnarray}\label{eqt: dps}
S_{Dp}^{\rm straight}=\frac{2\mathcal{N}U_0^{p+1-n}d^{3/2}}{g_sTd^{\frac{3}{4}(2n-p)}}\int_0^{z_T^{-3}} dz z^{-\frac{4}{3}}z^{-\frac{p+2n}{12}}\sqrt{\left(1+H^2d^3z\right)}z^{1/2}\ . 
\end{eqnarray}
The true vacuum is therefore determined by minimizing $\Delta S\sim
(S_{Dp}^{\rm curved}-S_{Dp}^{\rm straight})$; $\Delta S<0$ means
chiral symmetry breaking whereas $\Delta S>0$ means chiral symmetry
restoration. By looking at equation~(\ref{eqt: dpu}) and~(\ref{eqt:
  dps}) one can conclude that the phase diagram will depend on the
combination of $(p+2n)$ for an $n$-dimensional gauge theory with
flavour branes wrapping an internal $(p-n)$ directions. Specifically
speaking for $\{n=4,\, p=4\}$ and $\{n=3,\, p=6\}$ we should have
exactly the same phase diagram. It can be seen that the zeros of
$\Delta S$ depends on $H$ in exactly similar way for all these
examples, so the basic nature of the phase diagram would be universal.
All there is left for the exponent $(p+2n)$ then is to scale the
symmetry restoring temperature. Numerical results are shown below.

\begin{figure}[h!]
\begin{center}
\subfigure[] {\includegraphics[angle=0,
width=0.45\textwidth]{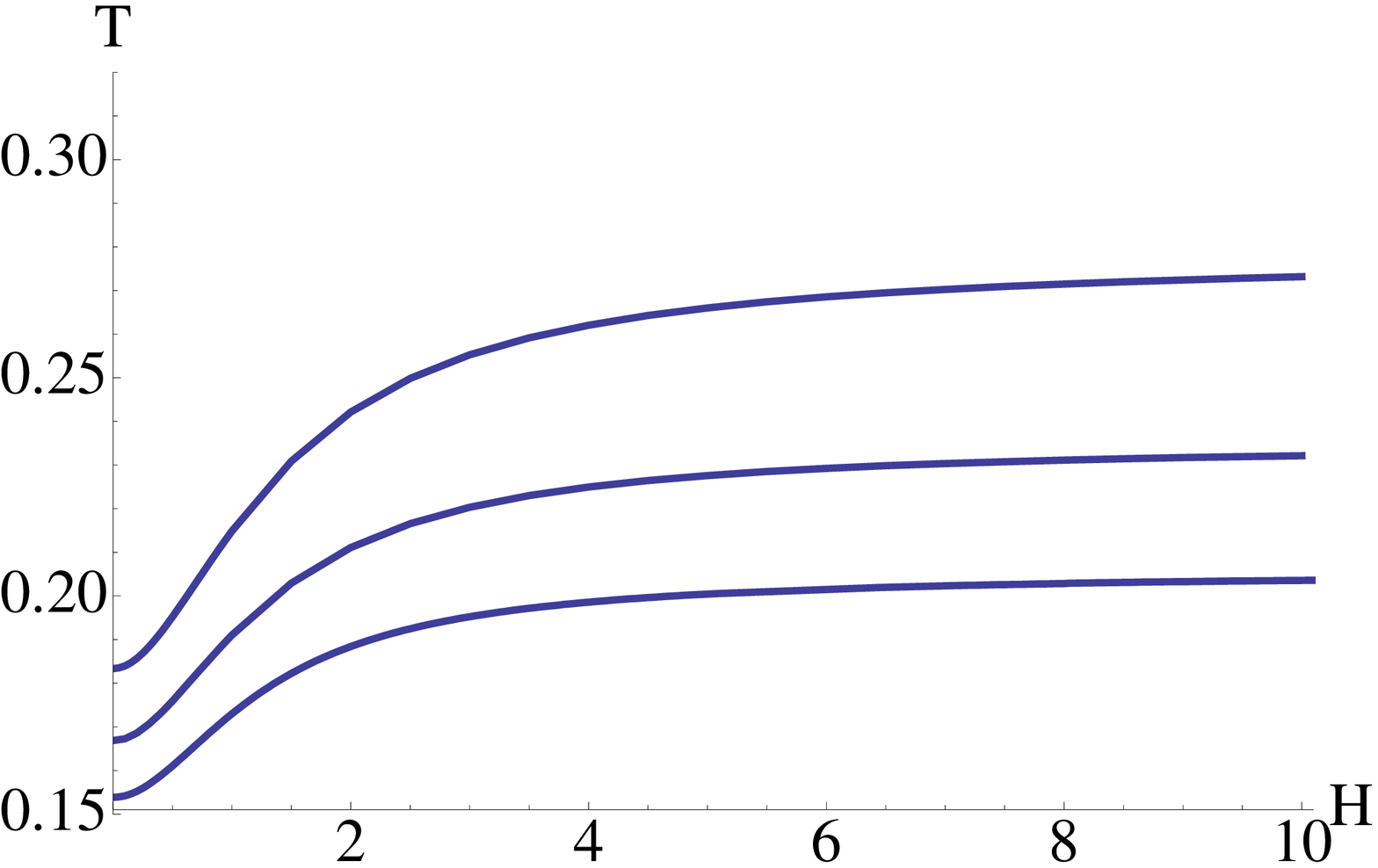} \label{fig: d4}}
\subfigure[] {\includegraphics[angle=0,
width=0.45\textwidth]{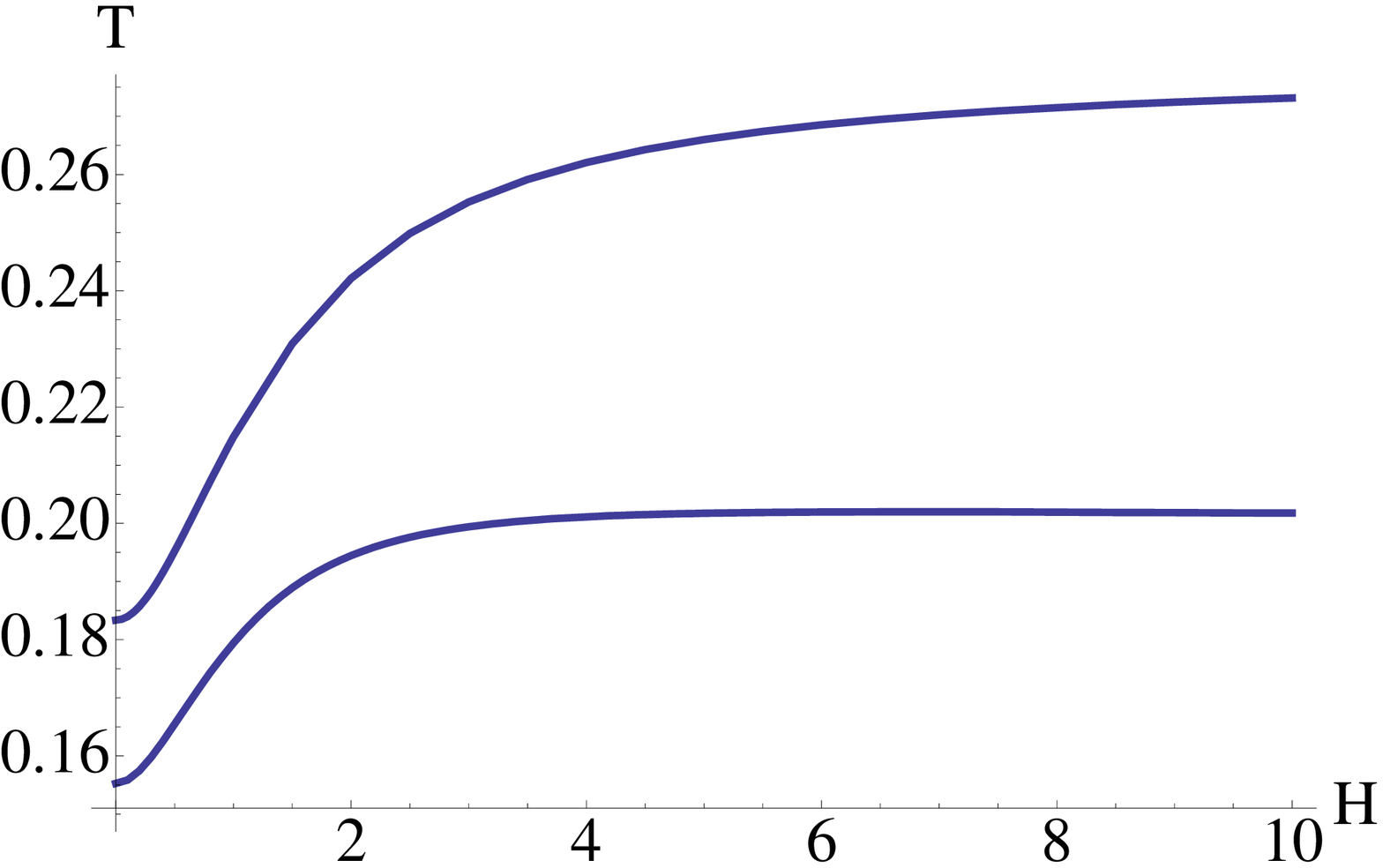} \label{fig: d3}}
\caption{\small The phase diagram for $n=4$ and $n=3$. Figure \ref{fig: d4} shows the phase diagram for the $4$-dimensional gauge theory, the curves correspond to $p=8,\, 6,\, 4$ from below to above respectively. Figure \ref{fig: d3} shows the corresponding phase diagram for $3$-dimensional gauge theory, where the curves correspond to $p=6,\, 4$ from below above respectively. We have set $d=1$.}
\end{center}
\end{figure}

Indeed we observe a generic nature of phase diagram in such
holographic models. We also find that the upper bound to symmetry
restoring temperature that can be reached introducing external
magnetic field depends on the dimension of the gauge theory and the
probe brane. We can observe that for a given dimension of the gauge
theory the lesser the number of directions wrapped by the probe brane
along the internal $S^4$, the higher is the symmetry restoring
temperature. Thus the information of the dimensions wrapped by the
probe brane is also encoded in the phase structure.

\section{A Note on Background Electric Field}

For this section we again consider the $8$-branes as flavours. We can
introduce a background electric field by considering the following
form of world--volume gauge field\cite{Karch:2007pd,Filev:2007gb}:
\begin{equation}
A_1(t,u)=(-Et+h(u))\ .\\
\end{equation}
This means we have a non-zero constant electric field along $x^1$. The
function $h(u)$ encodes the response of the fundamental flavours to
the external field, namely it encodes the information of the non-zero
current when flavours are free to move and therefore conduct. Note
that here we do not introduce any chemical potential, therefore in the
gauge theory there is no a priori candidate for carrying the charge.
However, there could still be current caused by pair creation in
presence of the electric field. We comment on some observations and
expectations henceforth.

For simplicity we assume $2\pi\alpha'=1$. We choose the same ansatz
for the probe brane profile as equation~(\ref{eqt: ansatz}). The DBI
action can be computed to be
\begin{eqnarray}\label{eqt: se}
S_{D8}&=& C\int du dt e^{-\phi} G_{xx}\left(detS_4\right)^{\frac{1}{2}}\left[G_{tt}G_{xx}g_{uu}+G_{tt}h'^2-g_{uu}E^2\right]^{1/2}\ , \nonumber\\
g_{uu}&=&G_{uu}+\tau'^2G_{\tau\tau}\ , 
\end{eqnarray}
where $C=(N_f \mu_8)/(V_{\mathbb{R}^3}V_{S_4})$, $G_{\mu\nu}$ is the
background metric and $g_{uu}$ is the induced metric component on the
world-volume of the $D8$-brane along $u$-direction.

Now we will have two constants of motion corresponding to the function
$\tau(u)$ and $h(u)$ as follows
\begin{eqnarray}\label{eqt: Eeom}
\frac{Ce^{-\phi}\sqrt{detS_4}G_{xx}(G_{tt}G_{xx}-E^2)G_{\tau\tau}\tau'}{\sqrt{G_{tt}h'^2-(E^2-G_{tt}G_{xx})(G_{uu}+G_{\tau\tau}\tau'^2)}}=B\ ,\nonumber
\end{eqnarray}
\begin{eqnarray}
\frac{Ce^{-\phi}\sqrt{detS_4}G_{xx} G_{tt}h'}{\sqrt{G_{tt}h'^2-(E^2-G_{tt}G_{xx})(G_{uu}+G_{\tau\tau}\tau'^2)}}=J\ .
\end{eqnarray}
The constants $B$ and $J$ are related to the minimum radial distance
along the world-volume of the probe brane ($U_0$ as before) and the
gauge theory current (a similar identification made in
ref.~\cite{Karch:2007pd} holds here also).

Solving equation~(\ref{eqt: Eeom}) numerically to look for all
possible solutions is a difficult problem. Nevertheless, it can be
shown from equation~(\ref{eqt: Eeom}) that for confined phase there
can be only the joined solution (curved as before), for which the
current vanishes if we impose the condition that the brane--anti-brane
pair join smoothly at $U_0$. This can be shown by formally solving
equation~(\ref{eqt: Eeom}) to obtain $\tau(u)$ and $h(u)$ as a
function of $u$ and the constants $B$ and $J$. If we expand the
solution around the joining point $u=U_0$ and demand that
$\tau'\to\infty$ as $u\to U_0$, we obtain that the current identically
vanishes. However, if the branes do join but not smoothly, this is no
longer necessarily true. One could, in this case, have cusp-like solutions for which the brane--anti-brane pair join at an angle and in order to stabilize the system it is necessary to consider the inclusion of a bunch of fundamental strings extending from the joining point $u=U_0$ to $u=U_{KK}$. This leads us to a construction much like in refs.~\cite{Rozali:2007rx, Bergman:2007wp, Davis:2007ka}, where the effect of baryons was considered. Therefore the symmetry broken phase may have a non-zero current carried by the baryons. The chiral symmetry is always broken in this
phase forced by the topology of the background. 

For the deconfined phase, it can also be shown in a similar way that
there exists curved solutions joining smoothly at some $U_0$, which
has zero current modulo the caveat mentioned in the last paragraph. This is intuitive from the gauge theory point of
view, since we are in a chiral symmetry broken phase therefore there is
no charge carrier present to conduct (ignoring the possibility of a baryon current). The possible effect of pair
creation is diminished by the existence of quark bound states in the
symmetry broken phase. However, for straight branes we expect non-zero
current to flow.

A familiar fact from studying flavours in electric field tells us that
the presence of electric field induces a so called ``vanishing locus"
for the probe DBI action. The ``healing" procedure (e.g., in
ref.~\cite{Filev:2007gb}) is to give a non-zero vev to the current. In
practice this is obtained by substituting the functions $\tau(u)$ and
$h(u)$ from equation~(\ref{eqt: Eeom}) in favour of the constants $B$
and $J$ in the action in equation~(\ref{eqt: se}) and demanding the
reality condition for the action for $u\in [U_T,\infty]$. The
condition leads to the following two equations
\begin{eqnarray}\label{eqt: cond}
&& \left(G_{tt}G_{xx}-E^2\right)^2=0\ ,\nonumber\\ 
 && B^2 e^{2\phi}G_{tt}-\left(G_{tt}G_{xx}-E^2\right)G_{\tau\tau}\left(C^2(detS_4)G_{tt}G_{xx}^2-e^{2\phi}J^2\right)=0\ . 
\end{eqnarray}
The two expressions are the terms in the on--shell action that go to
zero in the numerator and the denominator respectively. It can be
shown from equation~(\ref{eqt: cond}) that in order to have $J\not=0$
one has to have $B=0$, which corresponds to the straight brane
solutions.

For a given electric field, we can determine the position of the
vanishing locus $u_{eh}$ from the first condition in
equation~(\ref{eqt: cond}). Knowing $u_{eh}$ we can then extract the
current $J$, and therefore the conductivity using $J=\sigma E$. To
express the conductivity in terms of physical quantities let us recall
that
\begin{eqnarray}
\mu_8=(2\pi)^{-8}\alpha'^{-\frac{9}{2}}\ , \quad T=\frac{3\sqrt{U_T}}{4\pi R_{D4}^{3/2}}\ .\nonumber
\end{eqnarray}
Now we will restore the factors of $(2\pi\alpha')$ and also set
$R_{D4}^3=\pi\lambda\alpha'=1$. Combining everything we get the
following expression
\begin{eqnarray}
\sigma=\frac{4}{27}\lambda N_fN_c T^2\left(1+\frac{27}{32}\frac{E}{\lambda\pi^3T^3}\right)^{1/3}.
\end{eqnarray}
This conductivity is due to the melting of mesons at high temperature
and pair creation mediated by the electric field. Since we get this
formula using DBI action, it captures non-linear behaviour of the
conductivity with respect to the electric field.

The effect of the electric field on chiral symmetry breaking should be
to reduce the symmetry restoring temperature by polarizing the bound
states into constituent quarks. The analysis above points to the fact
that electric field works as expected from field theory perspective.
However, the energy consideration does not lead to the expected
result, because as electric field increases $u_{eh}$ also increases,
therefore the straight branes which extend all the way down to $U_T$
has more DBI action energy as compared to their curved counterparts,
which can only extend down to $u_{eh}$. We hope to address this issue in future.

\section{Conclusion}

We have extended the study of Sakai--Sugimoto model to include the
presence of external electric and magnetic fields, examining the
dynamics of the flavour sector in (an analogue of) the ``quenched''
approximation. We have seen that external magnetic field helps in
chiral symmetry breaking. This particular effect of an external
magnetic field has been referred to as magnetic catalysis in field
theory literature (see e.g., refs.~\cite{Miransky:2002eb}). Our
results and observations are consistent with results from those
approaches. We found that the chiral symmetry restoring temperature
increases with increasing magnetic field. We have further observed
that such holographic models have an upper bound for the symmetry
restoring temperature depending on the dimension of the gauge theory
and the probe brane.

We briefly studied the effect of external electric field, and though
we have not explored all of the details, we expect the dynamics to be
also consistent with the field theory intuition, although (as we did
for magnetic field here) the precise details should be interesting to
uncover. The presence of non--zero electrical conductivity is also an
avenue of further study. It would also be interesting to study Hall
effect in such models when both electric and magnetic fields are
present.

Another important avenue would be to study the meson spectra in the
presence of these external fields, where we expect to see effects such
as Zeeman splitting, the Stark effect and so forth (as observed in the
$D3/D7$ model in some of refs.~\cite{Filev:2007gb}). We leave these
and many other interesting aspects for future exploration.

\section*{Acknowledgments}

We would like to thank Tameem Albash, Nikolay Bobev, Veselin Filev and Ramakrishnan Iyer for discussions. This research was supported by the US Department of Energy.

\providecommand{\href}[2]{#2}\begingroup\raggedright


\begin{thebibliography}{10}

\bibitem{Maldacena:1997re}
Juan M. Maldacena, ``The large N limit of superconformal field theories and supergravity,''{\em Adv. Theor. Math. Phys.} {\bf 2} (1998) 231-252,
\href{http://www.arXiv.org/abs/hep-th/9711200}{{hep-th/9711200}}.



\bibitem{Witten:1998qj}
E. Witten, ``Anti-de Sitter space and holography,''{\em Adv. Theor. Math. Phys.} {\bf 2} (1998) 253-291,
\href{http://www.arXiv.org/abs/hep-th/9802150}{{hep-th/9802150}}.

\bibitem{Gubser:1998bc}
  S.~S.~Gubser, I.~R.~Klebanov and A.~M.~Polyakov,
  Phys.\ Lett.\  B {\bf 428}, 105 (1998)
  [arXiv:hep-th/9802109].

\bibitem{Aharony:1999ti}
O. Aharony, S.S. Gubser, J. Maldacena, H. Ooguri and Y. Oz ``Large N field theories, string theory and gravity,''{\em Phys. Rept.} {\bf 323} (2000) 183-386,
\href{http://www.arXiv.org/abs/hep-th/9905111}{{\tt hep-th/9905111}}.

\bibitem{Witten:1998zw}
E. Witten, ``Anti-de Sitter space, thermal phase transition, and confinement in  gauge theories,''{\em Adv. Theor. Math. Phys.} {\bf 2} (1998) 505-532,
\href{http://www.arXiv.org/abs/hep-th/9803131}{{hep-th/9803131}}.

\bibitem{Karch:2002sh}
A. Karch and E. Katz, ``Adding flavor to AdS/CFT,''{\em JHEP} {\bf 06} (2002) 043,
\href{http://www.arXiv.org/abs/hep-th/0205236}{{hep-th/0205236}}.

\bibitem{Sakai:2004cn}
T. ~Sakai and S. ~Sugimoto, ``Low energy hadron physics in holographic QCD,'' {\em Prog. Theor. Phys.} {\bf 113} (2005) 843-882,
\href{http://www.arXiv.org/abs/hep-th/0412141}{{hep-th/0412141}}.\\
T. ~Sakai and S. ~Sugimoto, ``More on a holographic dual of QCD,'' {\em Prog. Theor. Phys. } {\bf 114}
  (2006) 1083--1118,
\href{http://www.arXiv.org/abs/hep-th/0507073}{{hep-th/0507073}}.

\bibitem{Antonyan:2006vw}
E.~Antonyan, J.~A.~Harvey, S.~Jensen and D.~Kutasov,``NJL and QCD from string theory,''
\href{http://www.arXiv.org/abs/hep-th/0604017}{{hep-th/0604017}}.\\
E.~Antonyan, J.~A.~Harvey and D.~Kutasov,``The Gross-Neveu model from string theory,''
\href{http://www.arXiv.org/abs/hep-th/0608149}{{hep-th/0608149}}\\
E.~Antonyan, J.~A.~Harvey and D.~Kutasov,``Chiral symmetry breaking from intersecting D-branes,''
\href{http://www.arXiv.org/abs/hep-th/0608177}{{hep-th/0608177}}

\bibitem{Aharony:2006da}
O. ~Aharony, J. ~Sonnenschein and S. ~Yankielowicz, ``A holographic model of deconfinement and chiral symmetry
                  restoration,'' 
\href{http://www.arXiv.org/abs/hep-th/0604161}{{hep-th/0604161}}.\\

\bibitem{Casero:2005se}
  R.~Casero, A.~Paredes and J.~Sonnenschein,
  ``Fundamental matter, meson spectroscopy and non-critical string / gauge
  duality,''
  JHEP {\bf 0601}, 127 (2006)
  [arXiv:hep-th/0510110].\\ 
  K.~Peeters, J.~Sonnenschein and M.~Zamaklar,
  ``Holographic melting and related properties of mesons in a quark gluon
  plasma,''
  Phys.\ Rev.\  D {\bf 74}, 106008 (2006)
  [arXiv:hep-th/0606195].\\
  K.~Peeters, J.~Sonnenschein and M.~Zamaklar,
  ``Holographic decays of large-spin mesons,''
  JHEP {\bf 0602}, 009 (2006)
  [arXiv:hep-th/0511044].

\bibitem{Parnachev:2006dn}
A.~Parnachev and D.~A.~Sahakyan, ``Chiral phase transition from string theory,'' Phys. Rev. Lett.{\bf 97}, 111601 (2006),
\href{http://www.arXiv.org/abs/hep-th/0604173}{{ hep-th/0604173}}.\\
  A.~Parnachev and D.~A.~Sahakyan,
  ``Photoemission with chemical potential from QCD gravity dual,''
  Nucl.\ Phys.\  B {\bf 768}, 177 (2007)
  [arXiv:hep-th/0610247].

\bibitem{Horigome:2006xu}
N.~Horigome and Y. Tanii, ``Holographic chiral phase transition with chemical potential,'' JHEP {\bf 01}, 072 (2007),
\href{http://www.arXiv.org/abs/hep-th/0608198}{{hep-th/0608198}}.

\bibitem{Gepner:2006qy}
D. Gepner and Shesansu Sekahr Pal, ``Chiral symmetry breaking and restoration from holography,''
\href{http://www.arXiv.org/abs/hep-th/06108229}{{hep-th/0608229}}.

\bibitem{Rozali:2007rx}
  M.~Rozali, H.~H.~Shieh, M.~Van Raamsdonk and J.~Wu,
  ``Cold Nuclear Matter In Holographic QCD,''
  JHEP {\bf 0801}, 053 (2008)
   [arXiv:0708.1322 [hep-th]].
  
\bibitem{Bergman:2007wp}
  O.~Bergman, G.~Lifschytz and M.~Lippert,
  ``Holographic Nuclear Physics,''
  JHEP {\bf 0711}, 056 (2007)
  [arXiv:0708.0326 [hep-th]].
  
\bibitem{Davis:2007ka}
  J.~L.~Davis, M.~Gutperle, P.~Kraus and I.~Sachs,
  ``Stringy NJL and Gross-Neveu models at finite density and temperature,''
  JHEP {\bf 0710}, 049 (2007)
  [arXiv:0708.0589 [hep-th]].  
  
\bibitem{Aharony:2007uu}
  O.~Aharony, K.~Peeters, J.~Sonnenschein and M.~Zamaklar,
  ``Rho meson condensation at finite isospin chemical potential in a
  holographic model for QCD,''
  JHEP {\bf 0802}, 071 (2008)
  [arXiv:0709.3948 [hep-th]].
  
\bibitem{Parnachev:2007bc}
  A.~Parnachev,
  ``Holographic QCD with Isospin Chemical Potential,''
  JHEP {\bf 0802}, 062 (2008)
  [arXiv:0708.3170 [hep-th]].

\bibitem{Kruczenski:2003uq}
M.~Kruczenski, D.~Mateos, R.~C. Myers, and D.~J. Winters, ``Towards a
  holographic dual of large-N(c) QCD,'' {\em JHEP} {\bf 05} (2004) 041,
\href{http://www.arXiv.org/abs/hep-th/0311270}{{hep-th/0311270}}.

\bibitem{Babington:2003vm}
J.~Babington, J.~Erdmenger, N.~J. Evans, Z.~Guralnik, and I.~Kirsch, ``Chiral
  symmetry breaking and pions in non-supersymmetric gauge / gravity duals,''
  {\em Phys. Rev.} {\bf D69} (2004) 066007,
\href{http://www.arXiv.org/abs/hep-th/0306018}{{hep-th/0306018}}.

\bibitem{Kruczenski:2003be}
M.~Kruczenski, D.~Mateos, R.~C. Myers, and D.~J. Winters, ``Meson spectroscopy
  in AdS/CFT with flavour,'' {\em JHEP} {\bf 07} (2003) 049,
\href{http://www.arXiv.org/abs/hep-th/0304032}{{hep-th/0304032}}.

\bibitem{Albash:2006ew}
T. Albash, V. Filev, C.V. Johnson and A. Kundu, ``A topology-changing phase transition and the dynamics of flavour,''
\href{http://www.arXiv.org/abs/hep-th/0605088}{{hep-th/0605088}}.

\bibitem{Mateos:2006nu}
D.~Mateos, R.~C. Myers, and R.~M. Thomson, ``Holographic Phase Transitions with
  Fundamental Matter,''
\href{http://www.arXiv.org/abs/hep-th/0605046}{{hep-th/0605046}}.

\bibitem{Karch:2006bv}
A. Karch and A. O'Bannon, ``Chiral transition of N = 4 super Yang-Mills with flavor on a 3-sphere,''
\href{http://www.arXiv.org/abs/hep-th/0605120}{{hep-th/0605120}}.

\bibitem{Ghoroku:2005tf}
K.~Ghoroku, T.~Sakaguchi, N.~Uekusa, and M.~Yahiro, ``Flavor quark at high
  temperature from a holographic model,'' {\em Phys. Rev.} {\bf D71} (2005)
  106002,
\href{http://www.arXiv.org/abs/hep-th/0502088}{{hep-th/0502088}}.

\bibitem{Filev:2007gb}
V. G. ~Filev, C. V. ~Johnson, R. C. ~Rashkov and K. S. ~Viswanathan, ``Flavoured large N gauge theory in an external magnetic field,'' 
\href{http://www.arXiv.org/abs/hep-th/0701001}{{hep-th/0701001}}.\\
  T.~Albash, V.~G.~Filev, C.~V.~Johnson and A.~Kundu,
  ``Finite Temperature Large N Gauge Theory with Quarks in an External Magnetic
  Field,''
  arXiv:0709.1547 [hep-th].\\
  T.~Albash, V.~G.~Filev, C.~V.~Johnson and A.~Kundu,
  ``Quarks in an External Electric Field in Finite Temperature Large N Gauge
  Theory,''
  arXiv:0709.1554 [hep-th].\\
  J.~Erdmenger, R.~Meyer and J.~P.~Shock,
  ``AdS/CFT with Flavour in Electric and Magnetic Kalb-Ramond Fields,''
  JHEP {\bf 0712}, 091 (2007)
  [arXiv:0709.1551 [hep-th]].\\
  S.~Penati, M.~Pirrone and C.~Ratti,
  ``Mesons in marginally deformed AdS/CFT,''
  JHEP {\bf 0804}, 037 (2008)
  [arXiv:0710.4292 [hep-th]].
  
  
  \bibitem{Bergman:2007pm}
  O.~Bergman, S.~Seki and J.~Sonnenschein,
  ``Quark mass and condensate in HQCD,''
  JHEP {\bf 0712}, 037 (2007)
  [arXiv:0708.2839 [hep-th]].
  
 \bibitem{Bergman:2008sg}
  O.~Bergman, G.~Lifschytz and M.~Lippert,
  ``Response of Holographic QCD to Electric and Magnetic Fields,''
  arXiv:0802.3720 [hep-th]. 
  
  \bibitem{Karch:2007pd}
  A.~Karch and A.~O'Bannon,
  ``Metallic AdS/CFT,''
  JHEP {\bf 0709}, 024 (2007)
  [arXiv:0705.3870 [hep-th]].\\
  A.~O'Bannon,
  ``Hall Conductivity of Flavor Fields from AdS/CFT,''
  Phys.\ Rev.\  D {\bf 76}, 086007 (2007)
  [arXiv:0708.1994 [hep-th]].
  
  \bibitem{Miransky:2002eb}
  V.~A.~Miransky,
  ``Dynamics of QCD in a strong magnetic field,''
  arXiv:hep-ph/0208180.\\
  G.~W.~Semenoff, I.~A.~Shovkovy and L.~C.~R.~Wijewardhana,
  ``Universality and the magnetic catalysis of chiral symmetry breaking,''
  Phys.\ Rev.\  D {\bf 60}, 105024 (1999)
  [arXiv:hep-th/9905116].\\
  V.~P.~Gusynin, V.~A.~Miransky and I.~A.~Shovkovy,
  ``Dimensional reduction and catalysis of dynamical symmetry breaking by a
  magnetic field,''
  Nucl.\ Phys.\  B {\bf 462}, 249 (1996)
  [arXiv:hep-ph/9509320].
  
\end{thebibliography}
\end{document}